\documentclass[twocolumn]{aastex631}

\usepackage{xspace}
\usepackage{enumerate}

\defcitealias{Rey2020}{R20}
\newcommand{\yr}    	{\ifmmode \mathrm{yr} \else yr\fi}
\newcommand{\mpc}   	{\ifmmode \,\mathrm{Mpc}^{-3} \else \,Mpc$^{-3}$\fi}
\newcommand{\Msun}	    {\ifmmode \,\mathrm M_{\odot} \else $\,\mathrm M_{\odot}$\fi\xspace}
\newcommand{\Zsun}	    {\ifmmode \,\mathrm Z_{\odot} \else $\,\mathrm Z_{\odot}$\fi\xspace}
\newcommand{\Mhalo} 	{\ifmmode M_{\mathrm{halo}} \else $M_{\mathrm{halo}}$\fi\xspace}
\newcommand{\Rvir}  	{\ifmmode R_{200} \else $R_{200}$\fi\xspace}
\newcommand{\Mstar}	    {\ifmmode {M}_{\star} \else ${M}_{\star}$\fi\xspace}
\newcommand{\Mvir}	    {\ifmmode M_{\mathrm{halo}} \else $M_{\rm halo}$ \fi\xspace}
\newcommand{\Htwo}  	{\ifmmode {\rm H}_{2} \else ${\rm H}_{2}$ \fi}
\newcommand{\nH}    	{\ifmmode {n}_{\rm H} \else ${n}_{\rm H}$ \fi}
\newcommand{\cc}	    {\ifmmode {\rm cm}^{-3} \else ${\rm cm}^{-3}$ \fi}
\newcommand{\mperyr}	{\ifmmode \Msun{\rm yr}^{-1} \else $\Msun{\rm yr}^{-1}$ \fi}
\newcommand{\arepo}	    {{\small AREPO}\xspace}

\newcommand{\civ}	    {\ifmmode {\rm C}_{\rm IV} \else CIV \fi}
\newcommand{\mgii}	    {\ifmmode {\rm Mg}_{\rm II} \else MgII \fi}
\newcommand{\oi}	    {\ifmmode {\rm O}_{\rm I} \else OI \fi}
\newcommand{\perccm}	{\ifmmode {\rm cm}^{-2} \else ${\rm cm}^{-2}$ \fi}

\shorttitle{LYRA III: Reionization survivors}
\shortauthors{Gutcke et al.}
\graphicspath{{./}{figures/}}
\begin{document}

\title{LYRA III: The smallest Reionization survivors}

% \correspondingauthor{Thales A Gutcke, NASA Hubble fellow}
\email{thales@princeton.edu}
\author[0000-0001-6179-7701]{Thales A. Gutcke}\thanks{NASA Hubble Fellow}
\affiliation{Department of Astrophysical Sciences, Princeton University, 4 Ivy Lane, Princeton, NJ 08544, USA}

\author[0000-0002-7275-3998]{Christoph Pfrommer}
\affiliation{Leibniz-Institute for Astrophysics Potsdam (AIP), An der Sternwarte 16, 14482 Potsdam, Germany}

\author[0000-0003-2630-9228]{Greg L. Bryan}
\affiliation{Center for Computational Astrophysics, Flatiron Institute, 162 5th Ave, New York, NY 10010, USA}
\affiliation{Columbia University, Department of Astronomy, 550 W 120th St, New York, NY, 10025, USA}

\author[0000-0003-3308-2420]{R\"udiger Pakmor}
\affiliation{Max-Planck-Institut f\"ur Astrophysik, Karl-Schwarzschild-Str. 1, D-85748, Garching, Germany}

\author[0000-0001-5976-4599]{Volker Springel}
\affiliation{Max-Planck-Institut f\"ur Astrophysik, Karl-Schwarzschild-Str. 1, D-85748, Garching, Germany}

\author[0000-0002-7314-2558]{Thorsten Naab}
\affiliation{Max-Planck-Institut f\"ur Astrophysik, Karl-Schwarzschild-Str. 1, D-85748, Garching, Germany}

\begin{abstract}

The dividing line between galaxies that are quenched by reionization (``relics'') and galaxies that survive reionization (i.e. continue forming stars) is commonly discussed in terms of a halo mass threshold. We probe this threshold in a physically more complete and accurate way than has been possible to date, using five extremely high resolution ($M_\mathrm{target}=4~\Msun$) cosmological zoom-in simulations of dwarf galaxies within the halo mass range $1-4\times10^9~\Msun$. The employed LYRA simulation 
model features resolved interstellar medium physics and individual, resolved supernova explosions. In our results, we discover an interesting intermediate population of dwarf galaxies close to the threshold mass but 
which are neither full reionization relics nor full reionization survivors. These galaxies initially quench at the time of reionization but merely remain quiescent for $\sim500$~Myr.
At $z\sim5$ they recommence star formation in a synchronous way,  
 and remain star-forming until the present day. These results demonstrate that the halo mass at $z=0$ is not a good indicator of survival close to the threshold. 
 While the star formation histories we find are diverse, we show that they are directly related to the ability of a given halo to retain and cool gas. Whereas the latter is most strongly dependent on the mass (or virial temperature) of the host halo at the time of reionization, it also depends on its growth history, the UV background (and its decrease at late times) and the amount of metals retained within the halo.

\end{abstract}

\section{Introduction} \label{sec:intro}

Dwarf galaxies are observed to display a variety of star formation histories (SFHs, \citealt{Weisz2014}). Some SFHs show fairly continuous activity to the present day. Others are quenched at high redshift and then ``rejuvenated'' in the last few gigayears before the present day. Lastly, there are ``reionization relics'' that formed all their stars before the Universe was reionized ($z\sim8-7$) and have remained ``red and dead'' since.  To obtain  SFHs from observations, they can be reconstructed either from the color-magnitude diagram (CMD) or the spectral energy distribution (SED). This has been done for known dwarf galaxies in and around the Local Group, to a maximum distance of $\sim 4~\rm{Mpc}$ \citep{Weisz2014, McQuinn2010, Olsen2021}. 
\cite{Olsen2021} specifically find synchronized SF across their sample, requiring a large scale environmental cause or a general cosmological explanation. However, observations of such small systems are still only possible very close to the Local Group. Disentangling the contribution of reionization and environment is thus not easily possible for such faint systems. 

From a theoretical point of view, there should be a cut-off mass for galaxy formation due to the heating effect of the photoionizing background \citep{Rees1986, Efstathiou1992, Babul1992}.  This was simulated in detail by various authors \citep[e.g.][]{Thoul1996, Gnedin2000, Benson2002, Hoeft2006, Okamoto2008, Hambrick2009}, who find substantial gas mass loss and suppression of SF in low mass halos due the heating effect by the ultra-violet background (UVB). \cite{Okamoto2008} predict the characteristic, cut-off halo mass to be $M_c \sim 6.5 \times 10^9~\Msun$ at $z=0$. 

However, the characteristic mass is dependent on redshift and is lower at higher redshift.
\cite{Maccio2017} distinguish halos that remain dark or manage to form stars and propose that the ones that remain dark do so because their halos do not cross the characteristic mass at any time in their evolution.
\cite{Dijkstra2004} revisit the \cite{Thoul1996} model to show that dwarfs forming well before the peak of reionization ($z>10$) will be less affected, thus reducing the threshold mass for these objects.
This was modelled in detail by \cite{Benitez-Llambay2020}, who calculated the redshift--dependent critical halo mass for star formation before and after reionization. Assuming instantaneous reionization, they demonstrate a lower threshold mass before reionization, corresponding to the atomic hydrogen limit set by a virial temperature of $T_{200}\sim 7000~\rm{K}$. After reionization, the threshold mass is set by the temperature of ionized hydrogen: $2\times10^4~\rm{K}$. Once a halo is in thermal equilibrium with the background, the metallicity of the gas will not have much effect, since metal cooling lines are sub-dominant to hydrogen at this temperature. This critical mass theory connects the total mass growth history to the star formation history, since halos are only expected to form stars when they are more massive than the critical mass. If their growth does not keep pace with the critical mass growth, then they are predicted to cease star formation. Thus, we can call this a mass growth threshold.

These various models capture the general trends very well and have been substantiated by multiple galaxy formation simulations \citep[i.e.][]{Onorbe2015, PereiraWilson2022}. However, the evolution before reionization is unconstrained observationally, and theoretical predictions span three order of magnitude for the pre-reionization cut-off mass, covering the range $10^5-10^8~\Msun$. The lower end of this range is populated by models assuming that molecular hydrogen cooling leads to Population III star formation \citep{Machacek2001, Abel2002, Yoshida2003, Skinner2020, Kulkarni2021}. The upper end of the range is a result of assuming that atomic hydrogen is the dominant cooling channel, which happens when the virial temperature rises above $T_{200} = 7000~\mathrm{K}$ \citep[e.g.,][]{ Tegmark1997, Benitez-Llambay2020}. This translates to a mass threshold of around $10^8\,\Msun$ at $z=8$. 

However, since the critical mass is expected to be redshift--dependent and have a discontinuity at reionization, the assumed timing of reionization becomes a decisive parameter here as well. For example, the \cite{Tegmark1997} model allows for SF in halos of $\Mhalo=2\times10^6~\Msun$ at $z=30$. But this rises steeply to $10^8\Msun$ by $z=9$. New James Webb Space Telescope (JWST) detections of galaxies at $z\sim9-17$ show high rest-frame UV luminosities \citep{Harikane2022}. The authors conclude that this can either be explained by a higher than expected star formation rate surface density or a top-heavy stellar initial mass function (IMF). The former could indicate that the star formation efficiency is higher pre-reionization because it is not suppressed by a UV background. The latter could indicate that Population III stars are indeed more massive on average than later stars.

The aim of most of the models presented above is to predict the first primordial collapse threshold. This means that metal line and metal molecular cooling are not taken into account. So while this is an extremely interesting and important question, it really only applies for the very first, isolated and metal--free halos. However, the average $10^8\,\Msun$ halo at $z=10$ is already expected to be the product of the assembly (merging) of many smaller halos. Thus, the SFH we would like to understand is actually a cumulative history of the SF in each of these individual smaller halos. To complicate matters, the evolution and SF of such a halo at high redshift is quickly influenced by metal production \citep[i.e.][]{Curti2022}. 

After the first (metal--free) Population III (PopIII) stars form, it is expected that their metal production will enrich the surrounding gas and contribute to further cooling \citep[i.e.][]{Wise2012}. As shown in \cite{Jeon2017} and \cite{Gutcke2022}, SNe from star forming halos can eject metals beyond their own virial radii and enrich small neighboring halos that otherwise would not be able to cool efficiently via either molecular hydrogen or hydrogen line cooling. Once enriched, the cooling in these small systems is boosted by metal lines. This process allows certain halos in the vicinity of larger neighbors with masses down to $\lesssim10^3~\Msun$ to host stars prior to reionization. Thus, the assembly of a single $>10^8\,\Msun$ halo can include a combination of primordial and pre-enriched halos. 

\begin{figure*}[p]
    \centering
    \includegraphics[width=0.44\textwidth]{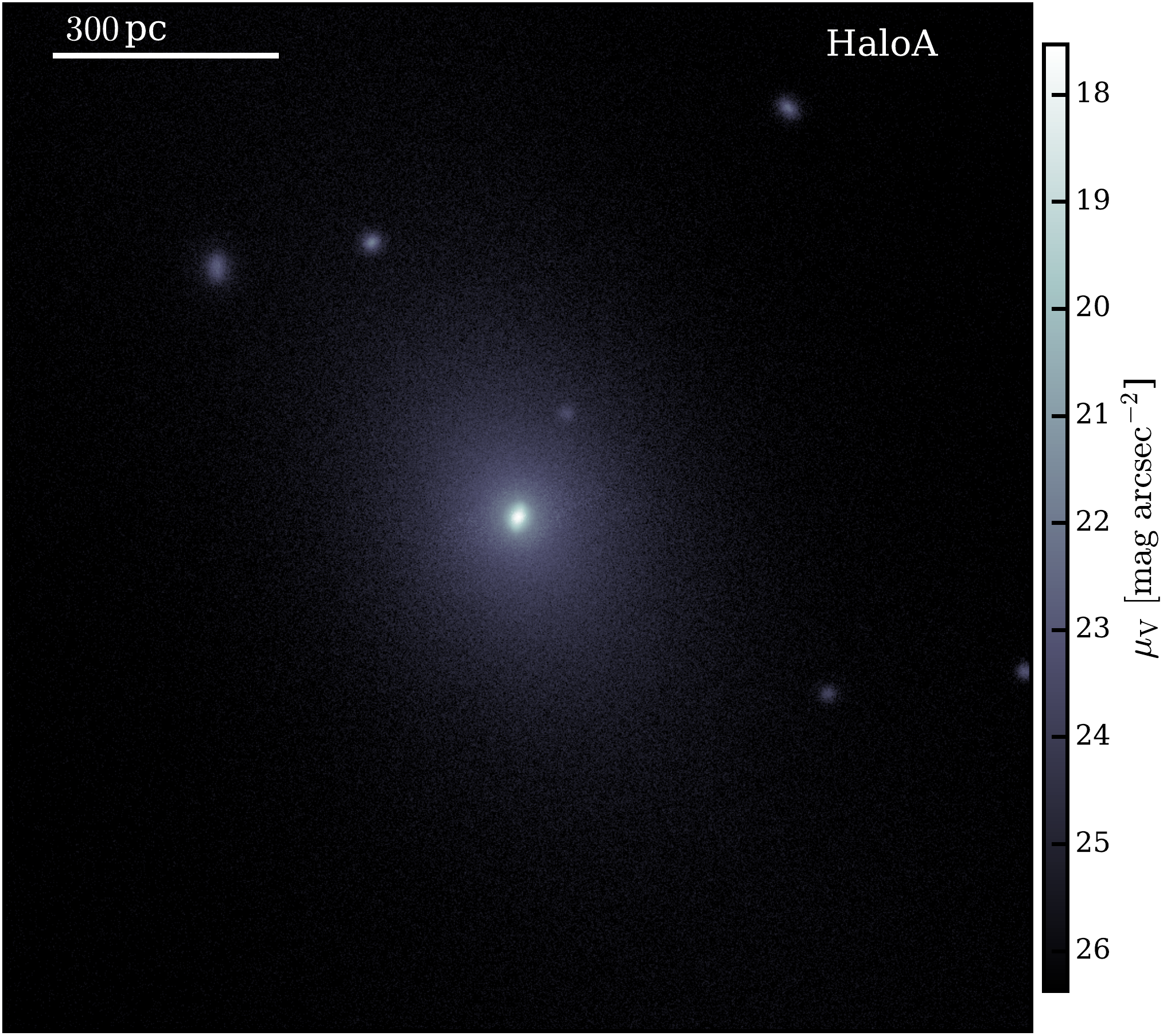}
    \includegraphics[width=0.45\textwidth]{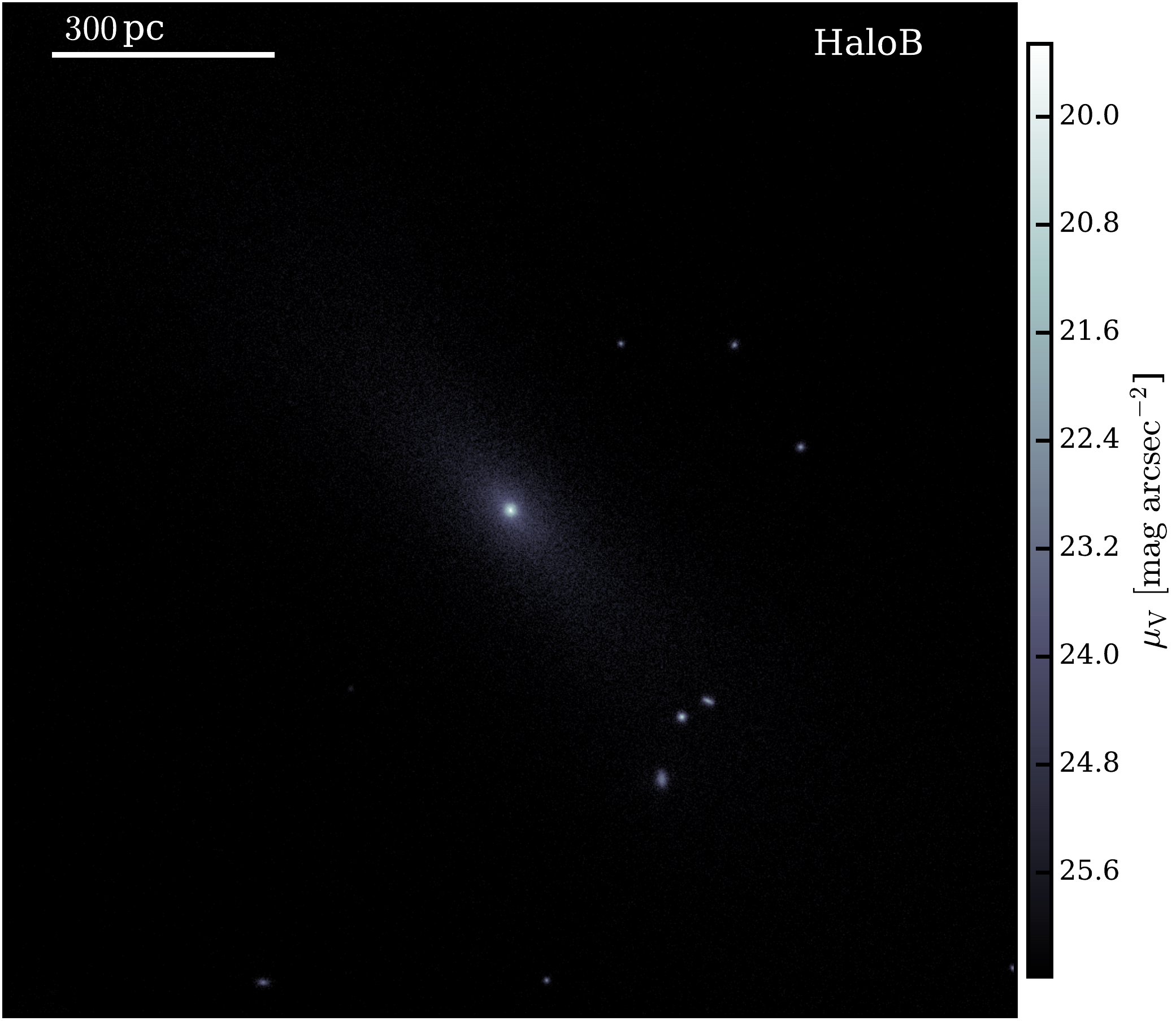}
    \includegraphics[width=0.45\textwidth]{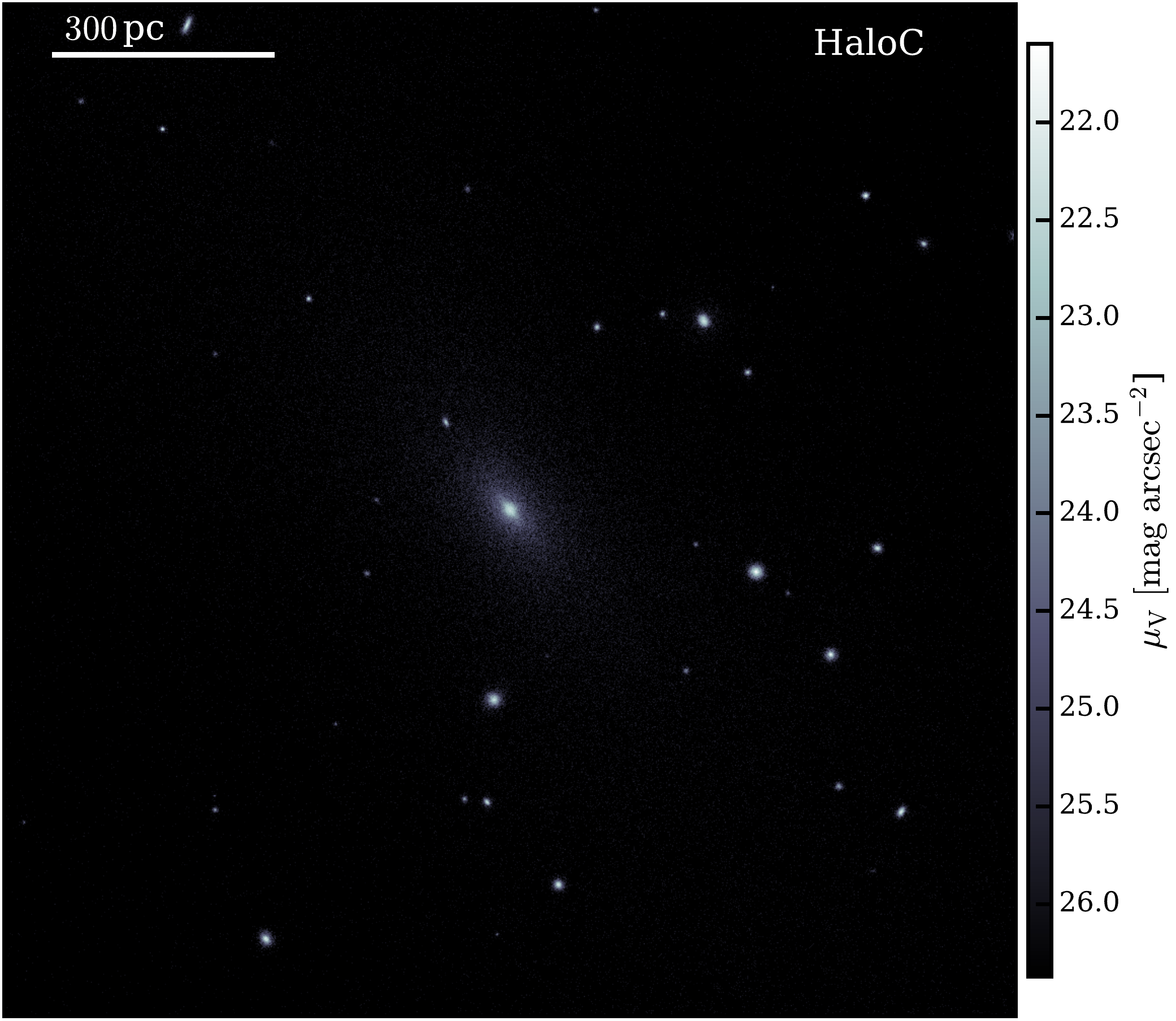}
    \includegraphics[width=0.45\textwidth]{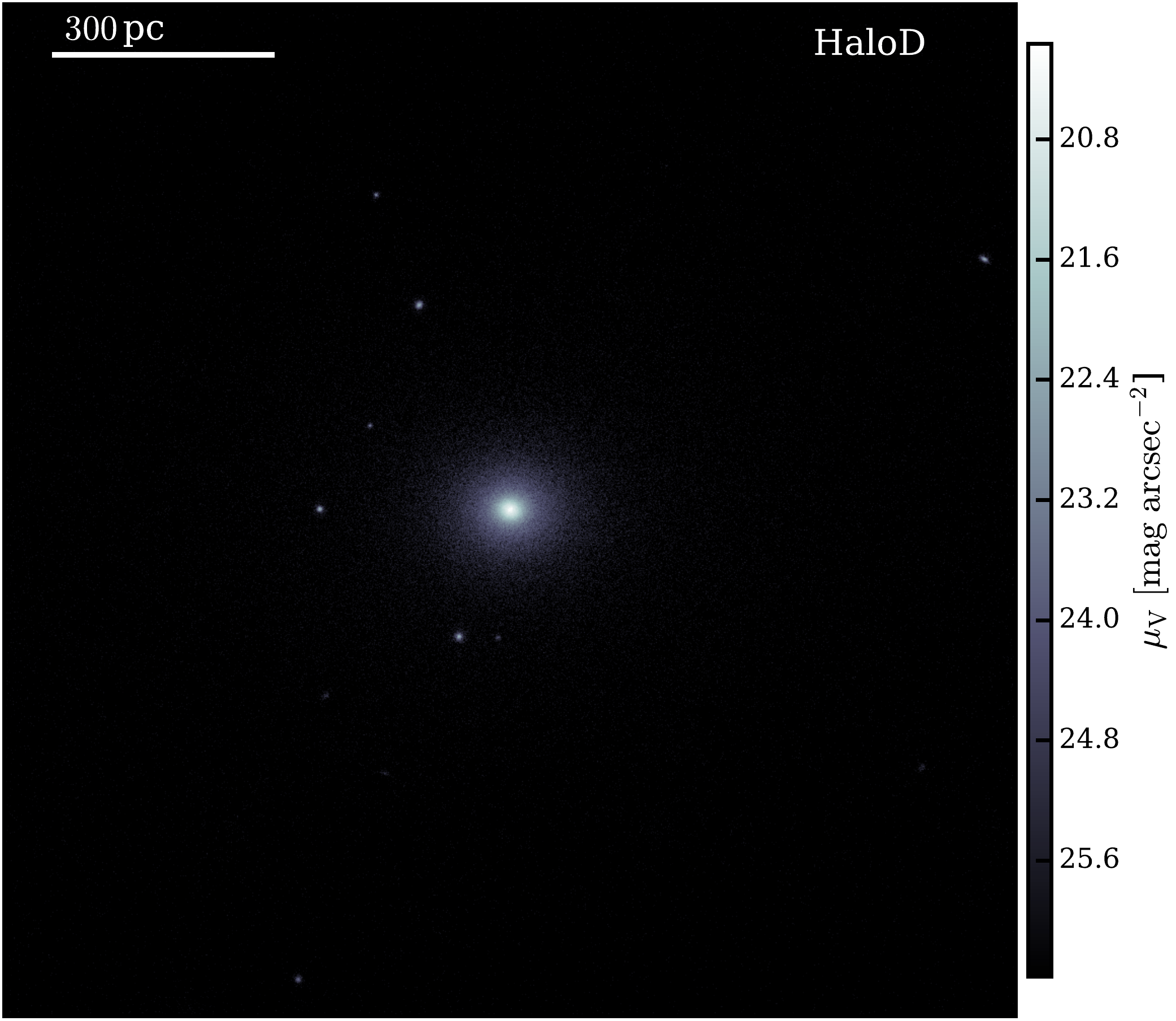}
    \includegraphics[width=0.45\textwidth]{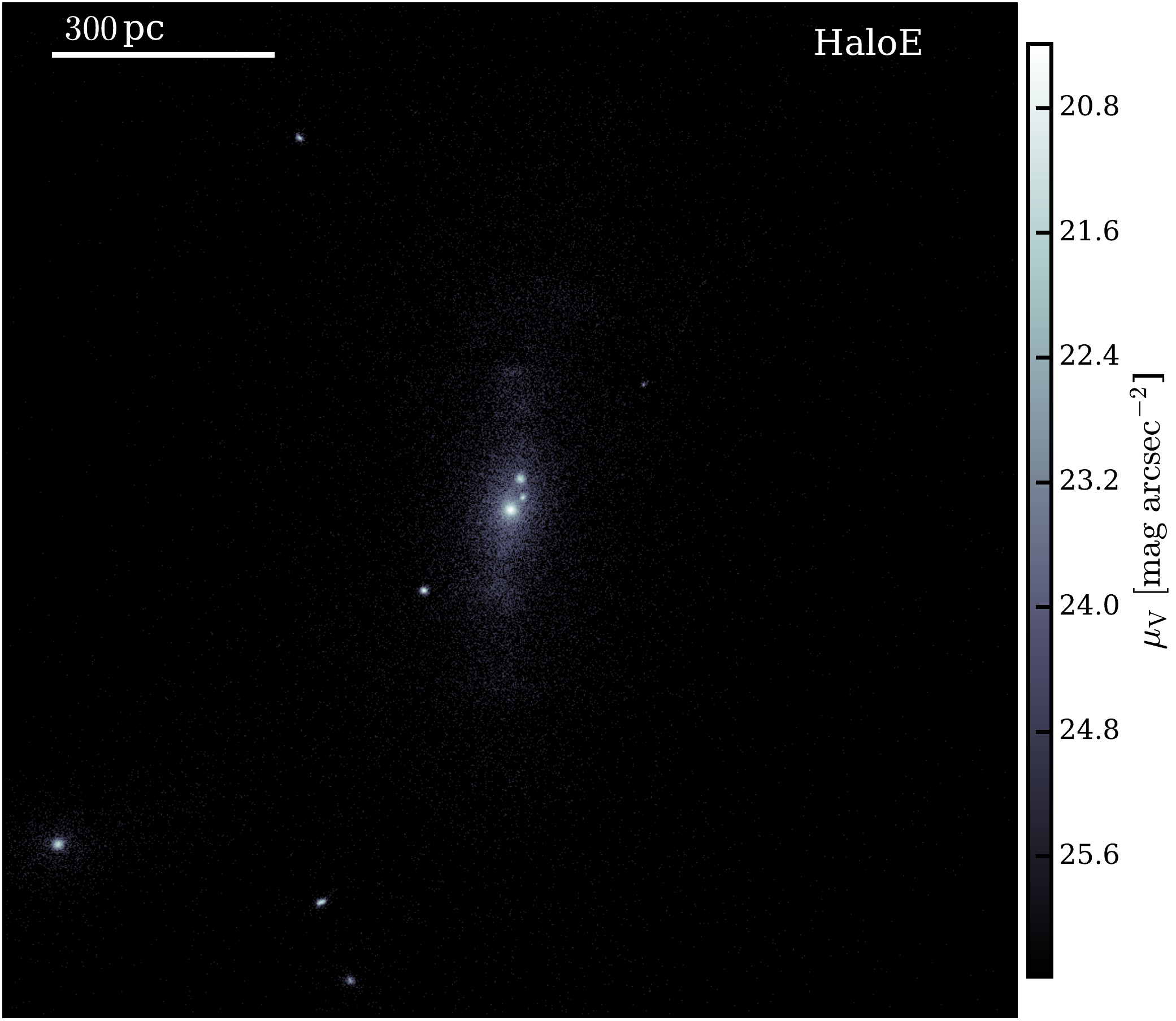}
    \caption{Mock V-band surface brightness maps for each simulation at $z=0$. Each image is $1.4~\mathrm{kpc}\times 1.4$ kpc. Notice the colorbar is rescaled for each image. The simulation is not rotated, only re-positioned to place the halo in the center.}
    \label{fig:SBmaps}
\end{figure*}

Lastly, as the halo mass increases and the UVB decreases, some or most of the metals can be retained within the halo. Combined, this creates favorable conditions for gas accretion, gas cooling and condensation. When the gas finally reaches the critical density for self-shielding, SF quickly reignites.

In this work, we aim to study the SFHs of the smallest dwarf galaxies that survive reionization and place their behavior into the theoretical context of the mass threshold models. While these models are generally valid, the detailed behavior of individual systems seems to be sensitive to various additional processes. In our analysis, we will attempt to disentangle the causes for the intermittent SF activity displayed by our simulations. Situations where multiple simulations demonstrate similar timing point to a cosmological cause that could be interpreted in terms of a larger-scale synchronicity. 

The paper is organized as follows: in Sec.~\ref{sec:model} we summarize the most relevant aspects of the simulation model. In Sec.~\ref{sec:SFH} we present the star formation histories of the five simulations. Then in Sec~\ref{sec:evolution} we discuss how these histories came about through an analysis of the halos' early evolution. In Sec.~\ref{sec:redshift0}, we compare the $z=0$ attributes of the simulations with observations of Local Group dwarf galaxies. Lastly, in Sec.~\ref{sec:discussion} and \ref{sec:conclusion} we discuss our results in the context of other simulation work and present our main findings. 

\setlength{\tabcolsep}{8pt}
\begin{table*}[t]
\centering
	\caption{Summary of simulation attributes: virial mass at $z=0$, stellar mass within $R_{200}$ at $z=0$, virial radius, stellar half mass radius, concentration parameter, bulk velocity of the halo, mean SFR until $z=4$, fraction of stars formed in the first 1~Gyr, V-band magnitude at $z=0$, luminosity at $z=0$, and mean stellar metallicity at $z=0$.}
		\begin{tabular}{l c c c c c c c c c c c}
		\hline \hline
			Name & $M_{200}$ & $M_\star$ & $R_{200}$ & $R_{1/2}$ & $c_\mathrm{max}$ & $v_\mathrm{bulk}$ & $\log \langle \dot M \rangle$ & $f_\mathrm{1Gyr}$ & $M_V$ & $\log L$ & $\log Z$ \\
			 & $[10^9 \mathrm{M}_\odot]$ & $[10^6 \mathrm{M}_\odot]$ & $[\mathrm{kpc}]$ & $[\mathrm{pc}]$ & & $[\mathrm{km~s}^{-1}]$ & $[\mathrm{M}_\odot ~\mathrm{yr}^{-1}]$ & $$ & $$ & $[\mathrm{L}_\odot]$ & $[\mathrm{Z}_\odot]$ \\
			 \hline
			HaloA & 3.34 & 9.42 & 28.7 & 311.5 & 13.5 & 177.5 & -2.3 & 0.519 & -11.5 & 6.69 & -1.02 \\
			HaloB & 2.82 & 1.26 & 29.8 & 483.2 & 12.7 & 146.6 & -3.2 & 0.992 & -8.8 & 5.79 & -2.60 \\
			HaloC & 2.16 & 1.05 & 27.2 & 878.4 & 13.0 & 46.3 & -3.1 & 1.000 & -8.6 & 5.57 & -2.22 \\
			HaloD & 1.72 & 1.59 & 25.3 & 119.6 & 24.8 & 120.0 & -2.8 & 0.830 & -9.0 & 5.65 & -1.47 \\
			HaloE & 0.29 & 0.22 & 7.2 & 107.0 & 14.7 & 11.0 & -3.7 & 1.000 & -8.4 & 5.55 & -2.52 \\
			\hline \hline
		\end{tabular}

	\label{tab:general}
\end{table*}

\section{Model \& Method} \label{sec:model}

In the current study, we present the first sample of five cosmological simulations run to $z=0$ using the LYRA galaxy formation model \citep{Gutcke2021, Gutcke2022}. LYRA is a comprehensive numerical model that includes a resolved ISM with a cooling prescription down to $10$K, individual (star-by-star) star formation, resolved supernovae and a subgrid model for PopIII star enrichment at high redshift. These prescriptions are implemented in the cosmological, hydrodynamical moving-mesh code \arepo \citep{Springel2010, Pakmor2016, Weinberger2020}. We direct the reader to these papers for a detailed description of the model and the code characteristcs. Below, we will summarize the most relevant aspects for this study.

The ultra-violet background (UVB) is implemented as an isotropic heating term following the rates presented in \cite{Faucher-Giguere2020}. HI reionization is set to $z=7.8$, but is $>90\%$ complete by $z=7.3$. HeII reionization is $>90\%$ complete at $z=3.0$. Gas is self-shielded from this heating following the fitting function presented in \cite{Rahmati2013}. In practice, the self-shielding factor depends on the gas density and the redshift. The majority of gas will be fully self-shielded above $\nH \gtrsim 10^{-2}~ \cc$. To enable the compatibility of heating and cooling rates, we choose to use the metal and low temperature cooling rates tabulated in \citet[][``UVB dust1 CR0 G0 shield1'']{Ploeckinger2020}. These were computed using \textsc{Cloudy} \citep{Ferland2017} and use the \cite{Faucher-Giguere2020} UVB as an input. Importantly, instead of the common ``1-zone'' model, these tables account for the full absorption of the incident radiation down to a fixed column density. We include these cooling rates from $z=20$ at the lowest value tabulated. The simulations are started with zero metallicity, thus no metal cooling can occur initially. Using the ``$10^6$'' case presented in \cite{Gutcke2022} as our fiducial model, we implement the enrichment by PopIII stars in a sub-grid manner. Running the halo finder \textsc{Subfind} \citep{Springel2001} on the fly, the metallicity of the gas within the virial radius is increased from $0$ to $10^{-4} \Zsun$ when a halo crosses the mass threshold of $M_\mathrm{PopIII}=10^6\Msun$. Once the gas metallicity is $10^{-4} \Zsun$ or higher, is it allowed to cool and form stars. 

Star formation follows a Schmidt relation \citep{Schmidt1959}:
\begin{equation}
    \dot M_\star = \varepsilon_\mathrm{SF}\frac{M_\mathrm{gas}}{t_\mathrm{ff}},
\end{equation}
where $\varepsilon_\mathrm{SF}$ is the star formation efficiency parameter and $t_\mathrm{ff}=\sqrt{3\pi/(32 G\rho)}$ is the free-fall time of the gas cell with density $\rho$, and $G$ is the gravitational constant.
Gas cells with $T/\mathrm{K} < 100$ and $3<\log(\nH/\cc)<4$ are eligible to form stars at an efficiency of $\varepsilon_\mathrm{SF}=2\%$. When $\log(\nH/\cc)>4$, the efficiency is increased to $\varepsilon_\mathrm{SF}=100\%$.

The initial conditions were taken from the EAGLE simulation \citep{Schaye2015} following the method presented in \cite{Jenkins2013}. The final zoom-in initial conditions simulate the entire EAGLE box ($L=100\,h^{-1}$~Mpc) at low resolution. Within this box, we define a high resolution region which encompasses the complete Lagrangian region of a single dwarf galaxy. Within this region gas can be refined and de-refined, cool radiatively and form stars. The DM mass resolution is $\sim80\,\Msun$, while the target gas cell mass is set to $4\,\Msun$. The gravitational softening length is $10$~pc for DM and $4$~pc for the gas and stars. The galaxies were chosen to fulfill an isolation criterion, namely to not interact with a larger galaxy for the duration of their lifetime.

\section{Results} \label{sec:results}

\subsection{Star formation histories} \label{sec:SFH}

In this work, we present a sample of five halos in the mass range $1-4\times 10^9~\Msun$. 
Fig.~\ref{fig:SBmaps} shows stellar surface brightness maps at $z=0$ for each of the five simulations. The general morphology is that of dwarf spheroidals (dSph), with a smooth extended distribution. However, there are two additional features of note in these mock images. Firstly, the center of the main spheroid seems to show signs of nucleation, a central stellar overdensity distinct from the extended profile. Nucleation is seen in dwarf galaxies in both the Fornax cluster \citep[Next Generation Fornax Survey]{Munoz2015, Eigenthaler2018} and the Virgo cluster \citep[Isaac Newton Telescope Virgo Survey]{Grant2005}.  Secondly, there is a clear indication of luminous substructure associated with each halo. We summarize the general attributes of the five simulations in Tab.~\ref{tab:general}.

In Fig.~\ref{fig:sfh} we present the five star formation histories as a function of redshift. More precisely, this is an initial--mass--weighted histogram of formation times of all stars within \Rvir at $z=0$. It includes accreted stars, but does not include stellar mass loss. The thin lines were computed using a bin width of $10$~Myr ($500$~Myr for the thick lines). The halos are ordered from top to bottom by halo mass at $z=0$. HaloA and HaloD show a similar SFH: they initially quench at $z=7.8$, recommence SF at $z\approx5$ and continue to form a significant fraction of their stars after reionization. Nevertheless, it should be pointed out that they are severely affected by reionization. Firstly, because they initially are quenched for approximately $400$-$500$~Myr before recommencing SF. Secondly, because the level of SF after reionization steadily decreases to values around $10^{-6} - 10^{-4}~\Msun~\yr^{-1}$, never again attaining the early (pre-reionization) value of $\sim10^{-2}~\Msun~\yr^{-1}$. 

HaloB is also initially quenched by reionization, but has a single burst around $z=5$, which reduces the central density such that no more stars are able form at that time. However, 6 billion years later the galaxy rejuvenates at around $z=0.3$ at a rate of $\sim10^{-5}~\Msun~ \yr^{-1}$. HaloC and HaloE are consistent with being reionization relics. Their star formation ceases at $z=7.8$, after which they are never again able to assemble sufficient gas to recommence forming stars. Interestingly, these two galaxies also form their first stars later than the other three galaxies. While HaloA, HaloB and HaloD begin forming stars shortly after $z=20$, these two halos do not host stars until around $z=16$. We also note that the $z=0$ halo mass is neither a good predictor of the final stellar mass, nor of the shape of the SFH.

\subsection{Why do certain halos survive Reionization?} \label{sec:evolution}

In the following, we will investigate the cause of the quenching at reionization in more detail. Most importantly, we will focus on the reasons why two (three, if we include HaloB) of the halos recommence their star forming activity, while the others do not. 

\setlength{\tabcolsep}{7pt}
\begin{table}
    \centering
    \begin{tabular}{c c c c c}
         \hline \hline
         Name & $z_\mathrm{rejuv}$ & $t_\mathrm{rejuv}~[\mathrm{Gyr}]$ & $\Delta t~[\mathrm{Myr}]$ & $t_\mathrm{sc}~[\mathrm{Myr}]$\\
         \hline
         HaloA & 5.14 & 1.140 & 470 & 155 \\
         HaloD & 5.04 & 1.168 & 493 & 163 \\
         HaloB & 4.59 & 1.312 & 640 & 128 \\
         \hline \hline
    \end{tabular}
    \caption{Initial quenching at $z_\mathrm{qu}\sim7.7$ or $t_\mathrm{qu}\sim0.67~\mathrm{Gyr}$ for all halos. Columns are: redshift of first rejuvenation after quenching, time of rejuvenation, duration of quenched episode, and mean sound crossing time of reionization--heated halo gas during the quenched episode.}
    \label{tab:sftimes}
\end{table}

\begin{figure*}
    \centering
    \includegraphics[height=0.675\textheight]{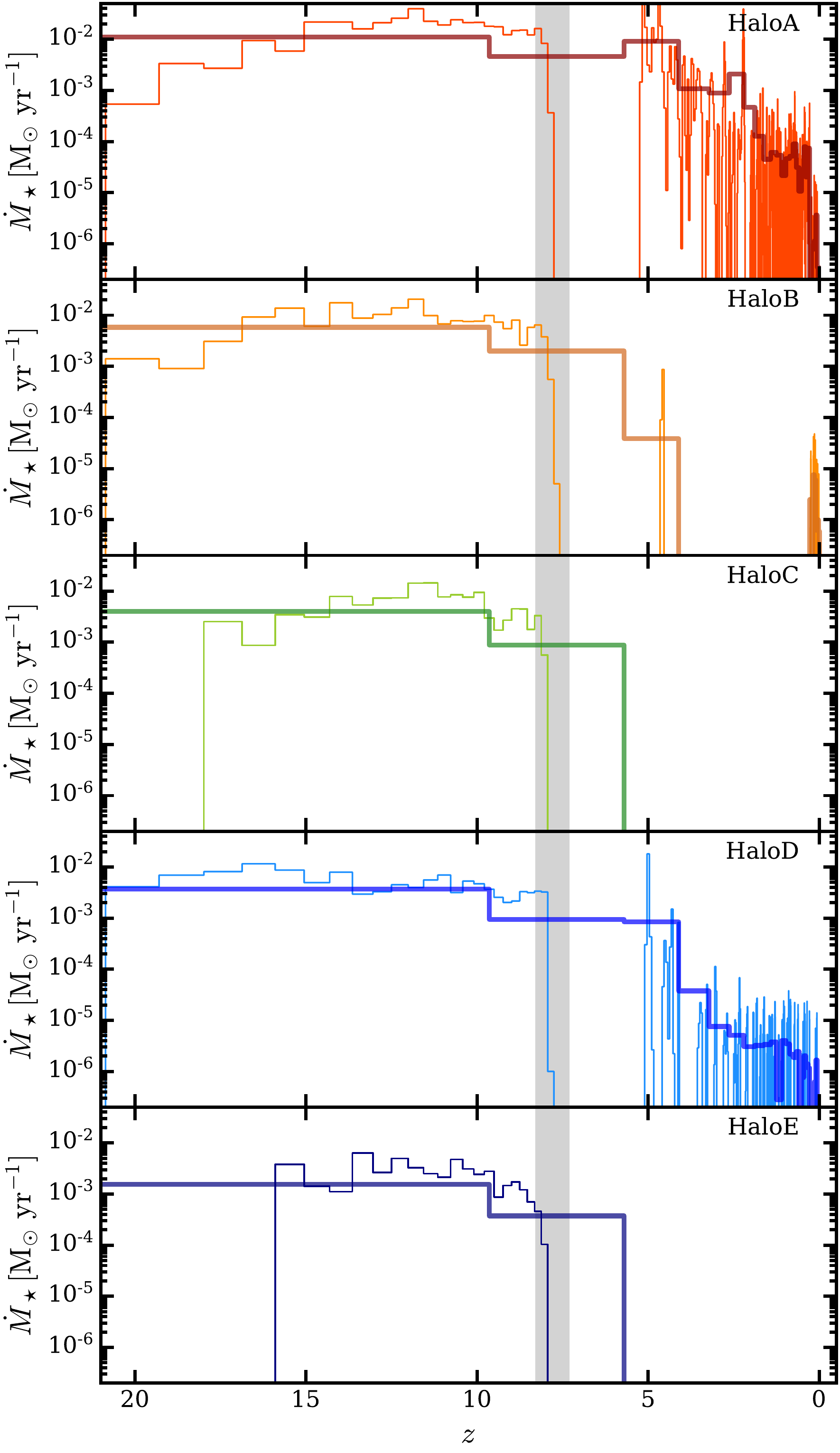}
    \includegraphics[height=0.7\textheight]{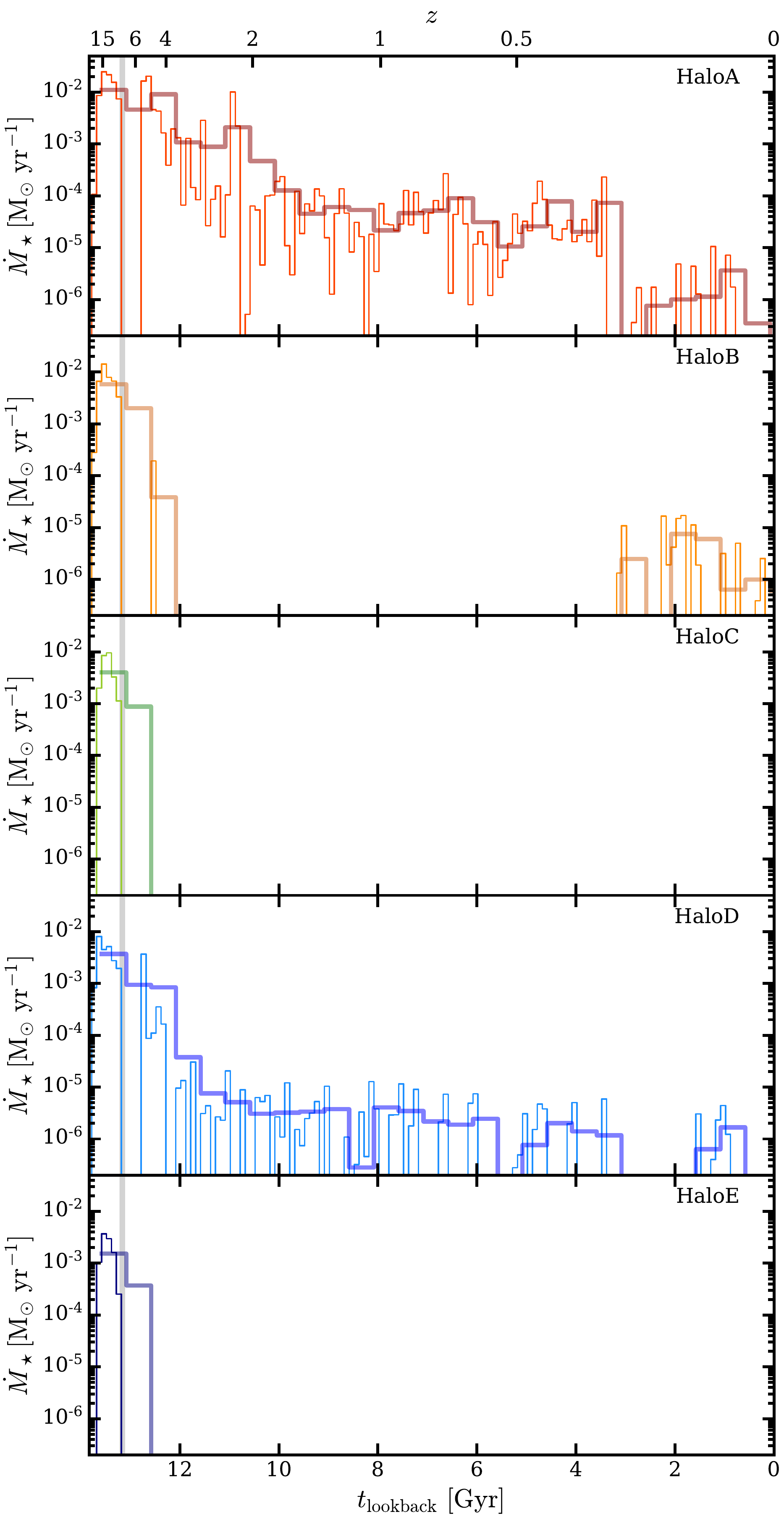}
    \caption{SFH in bins of 10 Myr (thin lines) and 500 Myr (thick lines). On the left the $x$-axis is redshift, while on the right it is lookback time in Gyr. HaloA and HaloD restart SF around $z=5$. HaloB and HaloE remain quenched after reionization. The grey bands in each panel indicate the time of hydrogen reionization.}
    \label{fig:sfh}
\end{figure*}

\begin{figure*}
    \centering
    \includegraphics[width=0.415\textwidth]{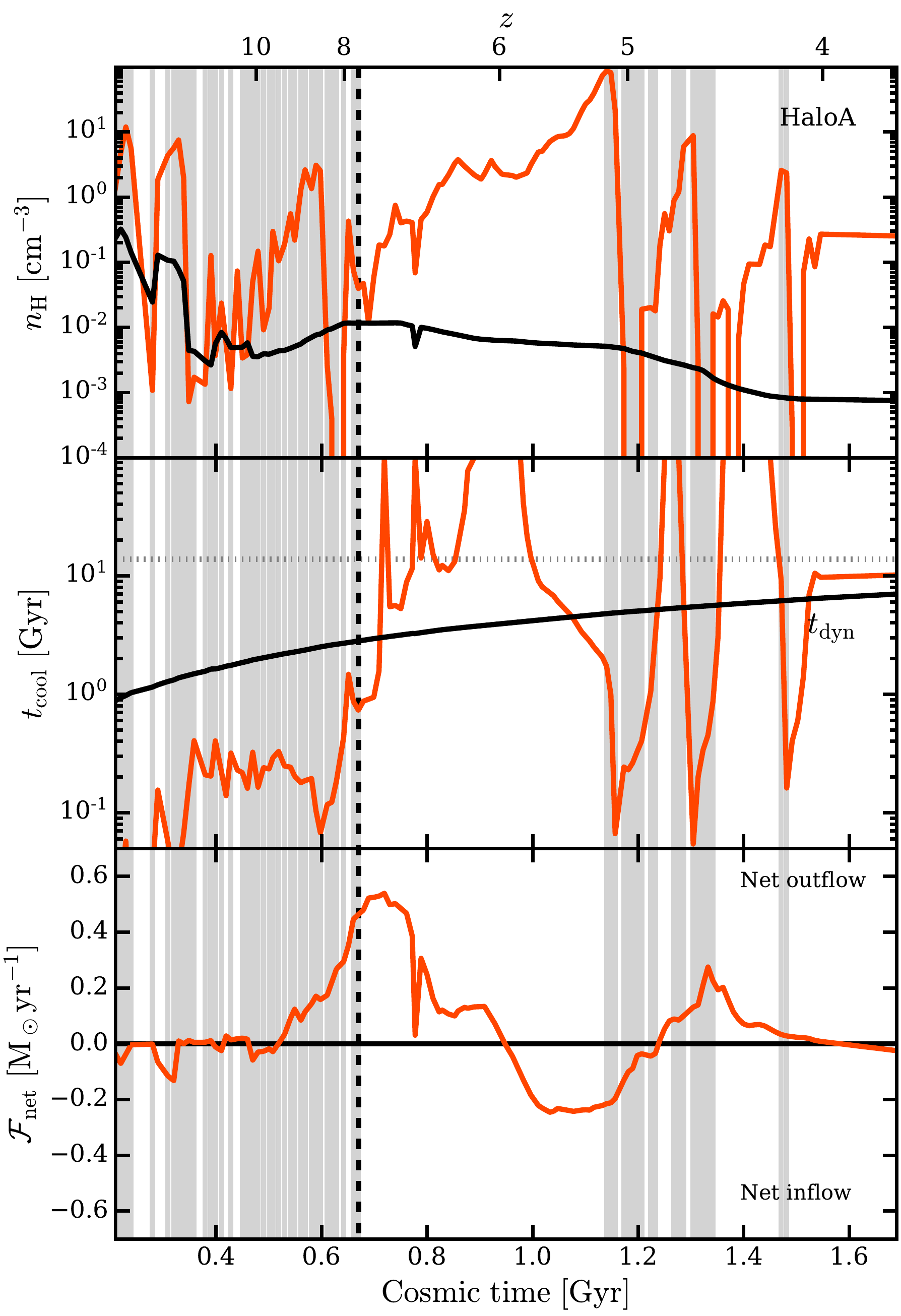}
    \includegraphics[width=0.42\textwidth]{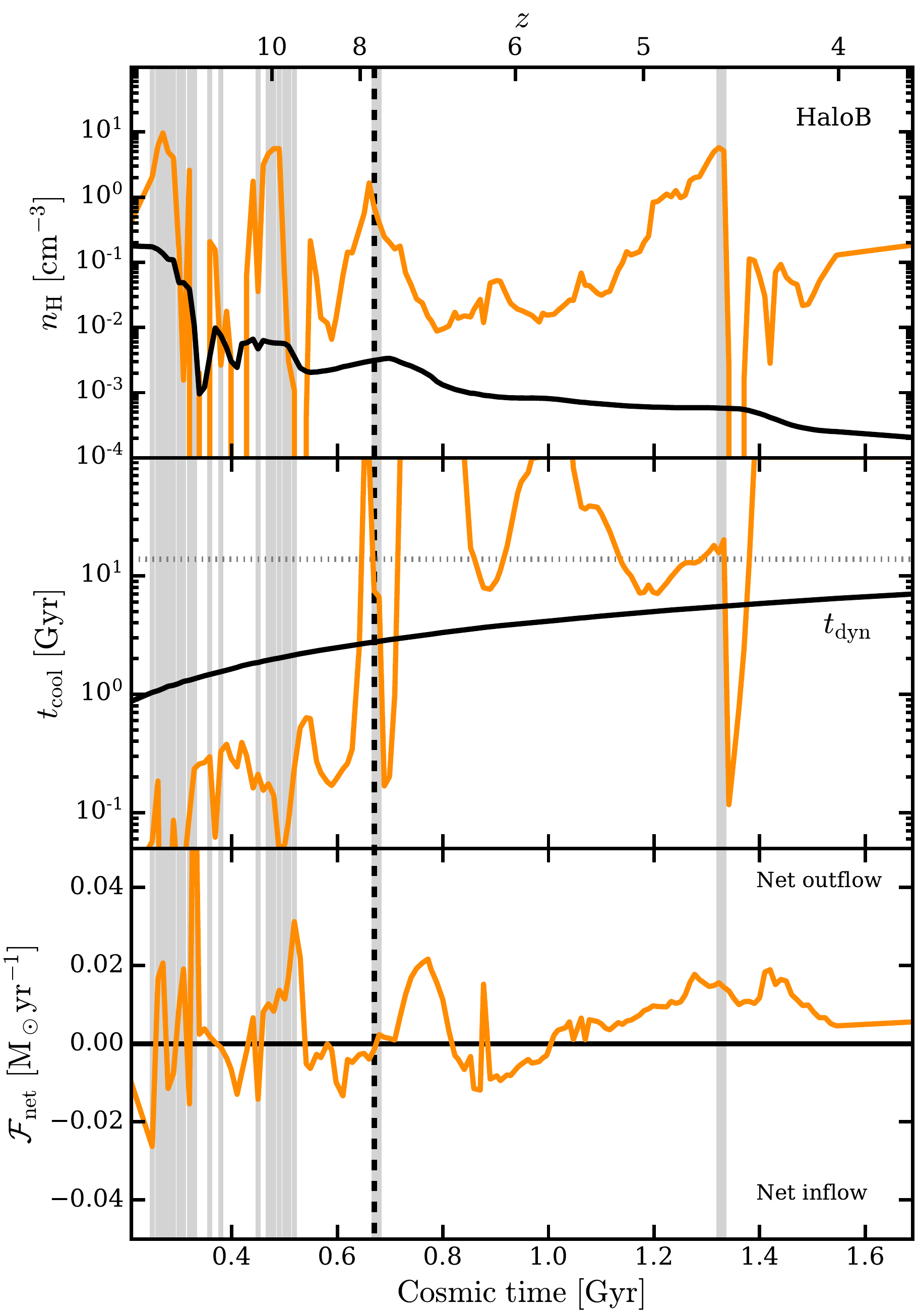}
    \includegraphics[width=0.438\textwidth]{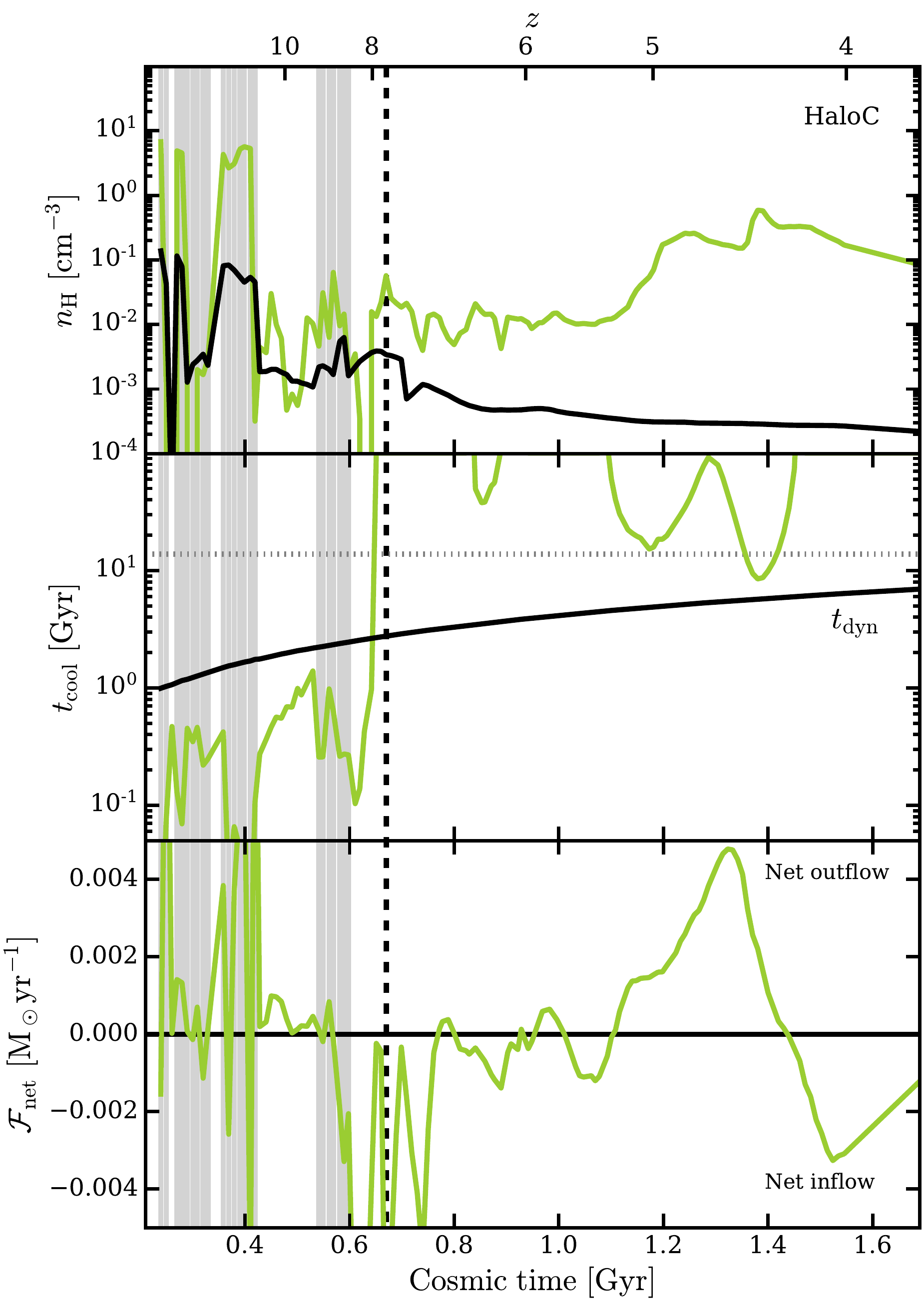}
    \includegraphics[width=0.435\textwidth]{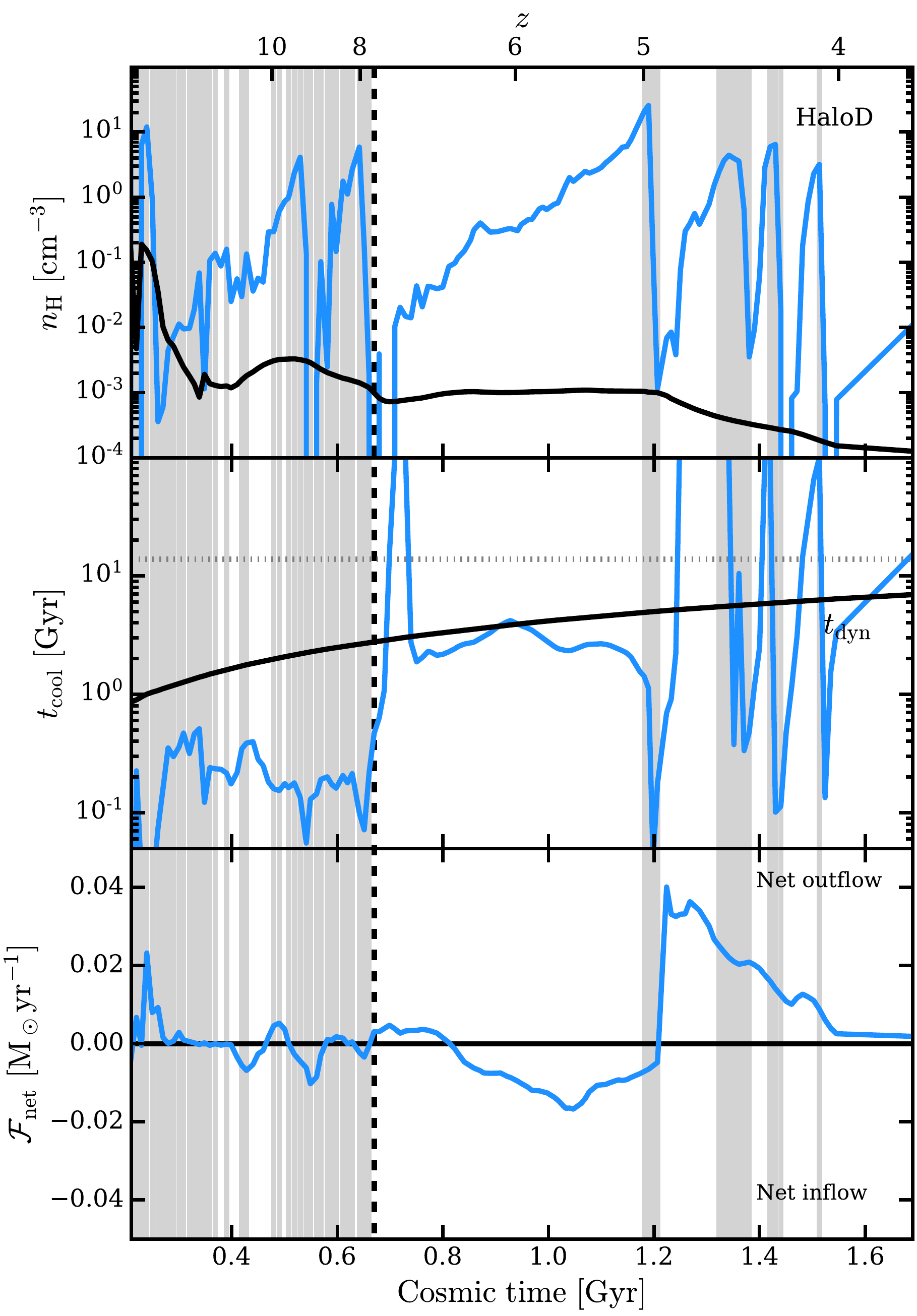}
    \caption{Each of the four figures shows three panels as a function of time for one simulation, respectively: the top panels show gas density within 40~pc (solid colored) and within \Rvir (black line). The central panels show the cooling time measured within the virial radius (colored), the dynamical time of the halo (black), and the Hubble time (dotted). The lower panels show the net mass flow rate (outflow rate minus inflow rate) across the spherical shell with radius \Rvir and width $0.1~\Rvir$. The grey bands in each panel indicate when SF is ongoing. The black vertical dashed line shows the time of quenching ($z=7.7$), which is just after the central reionization redshift ($z=7.8$). }
    \label{fig:flux}
\end{figure*}

Fig.~\ref{fig:flux} presents the evolutionary behavior of the four more massive simulations, HaloA-D, in the time interval $t = 0.2 - 1.7~\mathrm{Gyr}$. Each subplot shows data from one simulation. The top panel is the average hydrogen number density within the central $40$~pc (colored line) and within \Rvir (black line). The middle panel is the cooling time of the entire halo (colored line) computed as
\begin{equation}
    t_\mathrm{cool} =  \frac{u}{{\rm d}u/{\rm d}t},
\end{equation}
where $u$ is the total internal energy of gas cells within the halo. We also show the dynamical time of the halo (black line), which is
\begin{equation}
    t_\mathrm{dyn} =  \left(\frac{3 \pi }{32\, G \rho}\right)^{1/2},
\end{equation}
where $\rho$ is the average total mass density within the halo.
Finally, the bottom panel is the net gas flow rate across a spherical shell centered on \Rvir with a width of 0.1~\Rvir. The grey bands indicate ongoing SF within the halo at that time. We note that the grey bands differ slightly from the SFHs in Fig.~\ref{fig:sfh}, since the SFHs include all stars that reside within the main halo at $z=0$. Due to late accretion of luminous substructure, some of these stars are not yet present in the halo at higher redshifts.

Before reionization, all four simulations show similar behavior: the cooling time is well below the dynamical time and SF is fairly continuous. The central gas density varies but has peaks above $1\,\cc$. The outflow and inflow rates are generally well matched, meaning that SF is self-regulated: external accretion fuels SF and SNe drive outflows without preventing accretion. Thus, there is neither run--away cooling nor quenching. This self-regulation is strongly affected by the onset of reionization.

Hydrogen reionization begins at $z=8.3$ and completes at $z=7.3$, thus occurring at a mean redshift of $z=7.8$ (black dashed line). During this time, all gas that is not self-shielded is heated to approximately $1.5\times10^4~\mathrm{K}$. In response to this sudden heating, unshielded gas in the ISM and the entire circum-galactic medium (CGM) must expand. As \cite{Shapiro2004} show, even the self--shielded gas can eventually become photo--evaporated. The expansion causes outflows, so there is less net accretion onto the central region. SF comes to a sudden halt in all halos due to a lack of fuel. Cooling times of the gas increase above the dynamical time and even above the Hubble time (grey dotted line). The net flow rate may be outward due to the last SNe driving outflows into the now ionized medium, making the central gas density drop drastically.

Let us assume the majority of gas is in hydrostatic equilibrium before the onset of reionization. The gas is neither isothermal nor in thermal equilibrium with an external radiation field (since there is none in this model). A pressure (and temperature) gradient across the halo supports the non-SF gas. We can approximate reionization as an instantaneous (non-adiabatic) heating process.

The sudden heating of the gas by reionization raises the pressure. To reestablish equilibrium, the gas in the halo will expand on the sound crossing time scale, which is the time it takes to cross the halo when travelling at the local sound speed. It is straightforward to compute the adiabatic sound speed assuming an ideal gas:
\begin{equation}
    c_s = \sqrt{\gamma \frac{k_\mathrm{B}T}{\mu m_p}} \approx 18.4~\mathrm{km~s}^{-1} \left(\frac{T}{1.5\times10^4~\mathrm{K}}\right)^{1/2},
\end{equation}
where $\gamma=5/3$ is the adiabatic index, $k_\mathrm{B}$ is the Boltzmann factor, $T$ is the temperature, $m_p$ is the proton mass, and we assume a mean molecular weight of ionized hydrogen and singly ionized helium after reionization, $\mu=2/(3X+1)\approx0.61$, where $X=0.76$ is the primordial hydrogen mass fraction. The two halos with early rejuvenation, HaloA and HaloD undergo the same quenching and recommencement of SF at approximately the same time. They have an average virial radius of $\Rvir\approx3~\mathrm{kpc}$ between $z=7.8 - 5$. Thus, the sound crossing time of these halos is $t_\mathrm{sc}\approx160$~Myr.

Due to the sudden heating and subsequent expansion, some gas may become unbound, as seen in the net outflow rates. However, most of the gas will cool and re-collapse due to gravity. The net flow rates reverse and inflow begins to dominate after around one sound crossing time, driving gas back inside the virial radius. The initial expansions and subsequent recollapse can be seen as oscillation (or ``breathing'') of the halo in response to reionization. 

The total gas density (black line) in HaloB and HaloC steadily decreases immediately after reionization. There is photo-evaporation of the gas at and beyond the virial radius. Accretion is halted and outflows dominate. 
However, within one sound crossing time all four halos reestablish a central gas reservoir. The minimum density required to do this is the self-shielding threshold of  $\sim10^{-2}~\cc$. After that, the central gas densities (colored lines) either remain constant or rise steadily. This accretion occurs despite the total halo density (black lines) remaining constant or decreasing. This implies that most gas does not escape the halo, but is initially suspended in the CGM until it is slowly re-accreted onto the central zone. 

As the central densities increase, the cooling time decreases. Once it drops below the dynamical time again, SF can start again. This happens after $\sim$500~Myr in HaloA, HaloB and HaloD. The time required for this is a combination of the sound crossing time and the time required for outwards angular momentum transport and radiative cooling. We suggest that this $\sim$500~Myr break results from intrinsic properties of these halos that are at the edge between quenching and survival. The fact that all three halos display the same behavior can be interpreted as synchronicity with a cosmological (reionization) cause.

\begin{figure}
    \centering
    \includegraphics[width=\columnwidth]{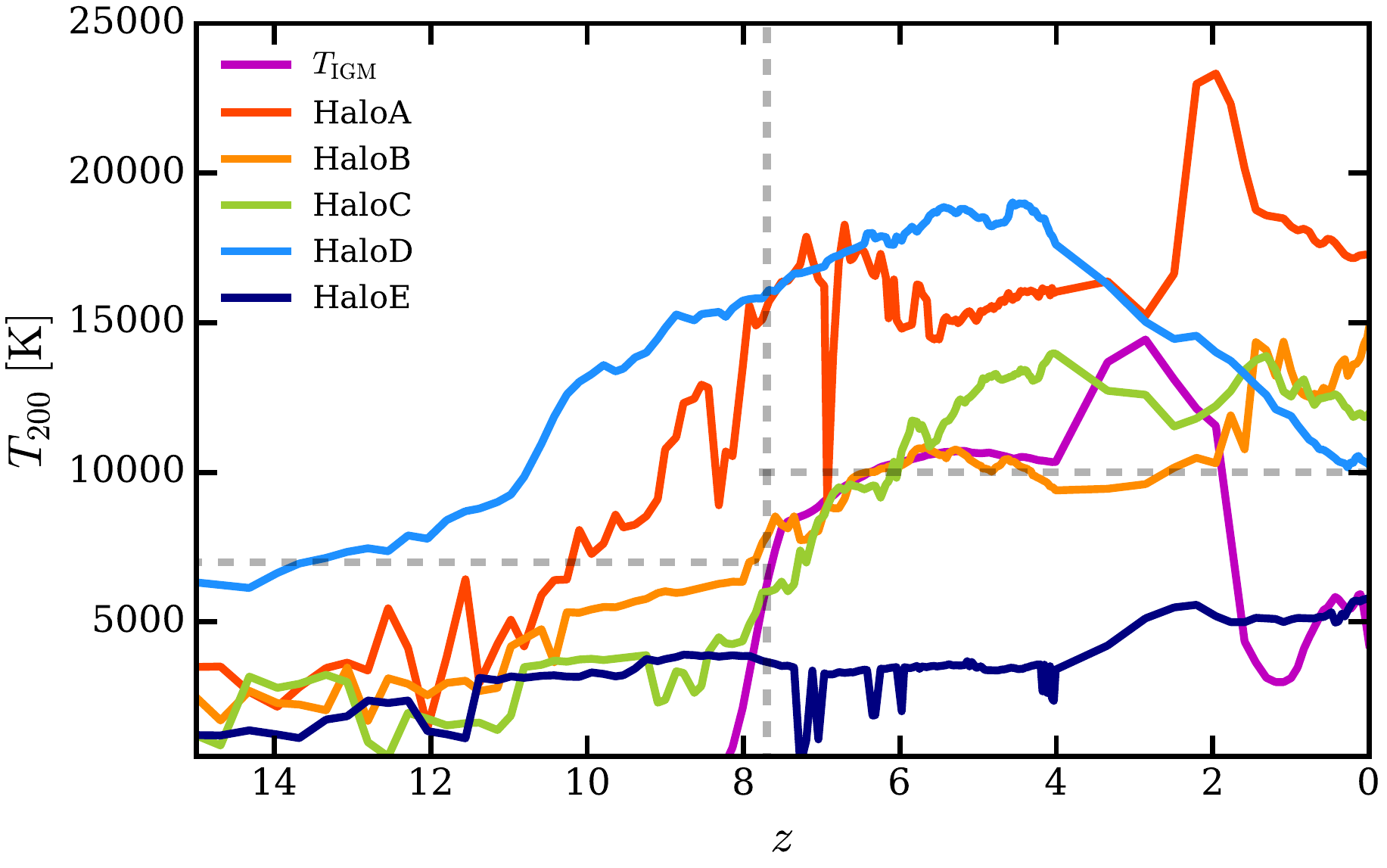}
    \includegraphics[width=0.98\columnwidth]{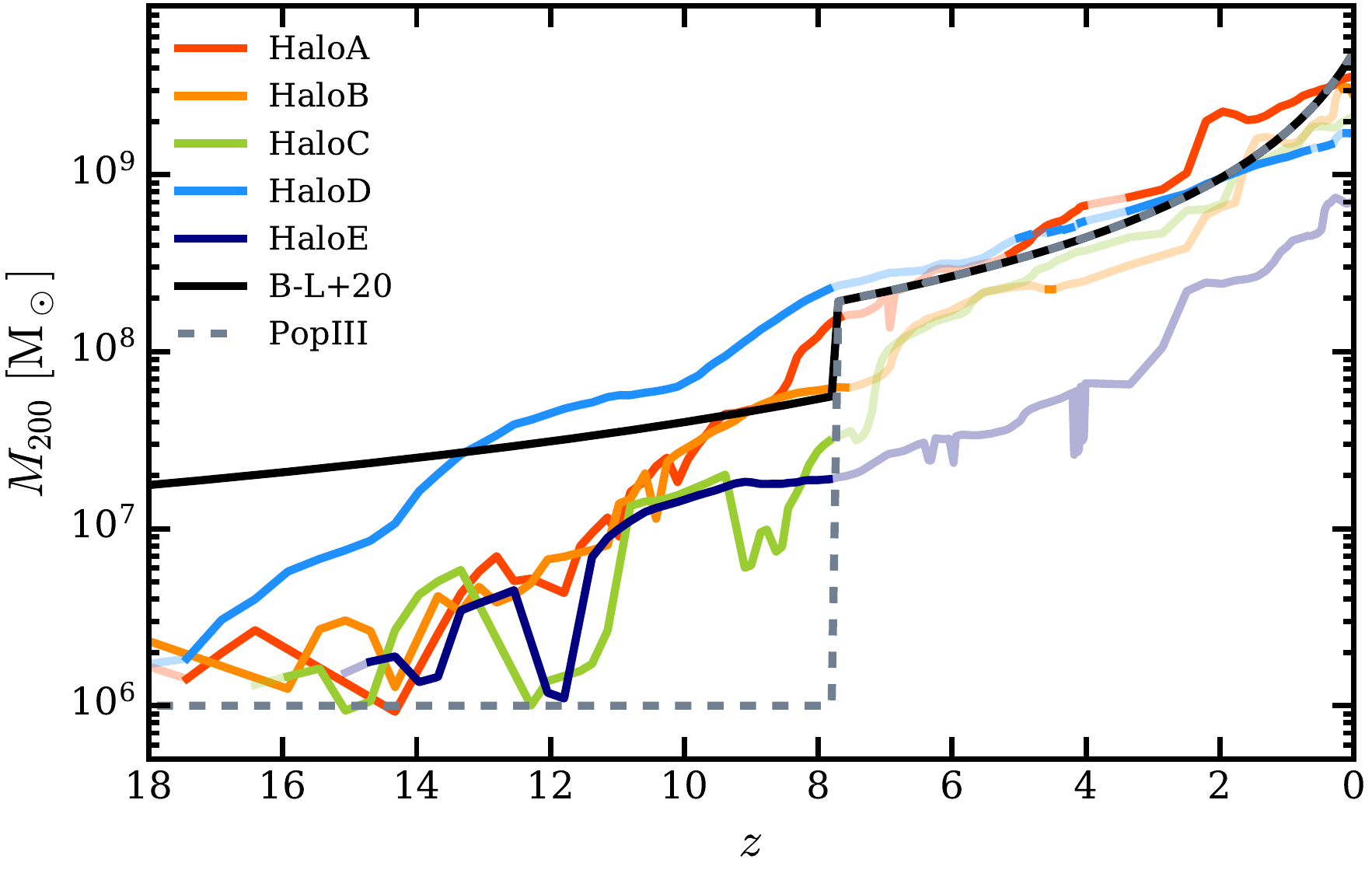}
    \caption{\textit{Top:} Evolution with redshift of the virial temperature of the halos. The horizontal dashed grey line shows $T=10^4\,\mathrm{K}$, the limit for atomic cooling in halos. The vertical dashed grey line shows the approximate timing of reionization when all halos halt their star formation. For comparison, we also plot the IGM temperature. \textit{Bottom:} Evolution with redshift of the virial mass of the halos. The opacity of the lines indicate when SF is ongoing (full) or quenched (opaque). For comparison, we plot the model of \cite{Benitez-Llambay2020} as a solid black line. The horizontal grey dashed line indicates the PopIII enrichment threshold in our model (before reionization) while we show the atomic cooling threshold of $10^4$ K for $z<7.7$. The vertical grey dashed line denotes the quenching redshift of $z=7.7$.}
    \label{fig:T200}
\end{figure}

HaloB shows a slightly different behavior than HaloA and HaloD after reionization. Already before reionization it shows a more bursty and sporadic SF with periods of lower gas densities. There is a final burst of SF at the time of reionization and some subsequent net outflow. However, the net flow does not turn around to become an inflow in the same way as in the other two halos. Instead, there is a low but persistent net outflow, meaning that the halo is photo-evaporating at its boundary. So while we can see a slow increase in the central gas density, this is material from the CGM, not new accretion. The central density reaches a sufficiently high value that SF recommences but is interrupted after a single burst, sending the cooling time to values greater than the Hubble time. SF is quenched for the next 6~Gyr. Then, at $z\approx 0.3$, it recommences at a low rate. 

Lastly, we will look at HaloC (which is very similar to HaloE and not shown here for simplicity). SF is quenched slightly before the peak of reionization, showing that the halo is especially sensitive to the external radiation field. The cooling time increases rapidly and the net flow rates are very small (note the $y$-axis is rescaled), showing no large-scale accretion. Interestingly, while this halo does not manage to form stars again, there is nevertheless an indication of activity around the same time that the other halos rejuvenate. The cooling time decreases below the Hubble time and the gas density rises slightly. After this, the activity ceases.

\begin{figure}
    \centering
    \includegraphics[width=\columnwidth]{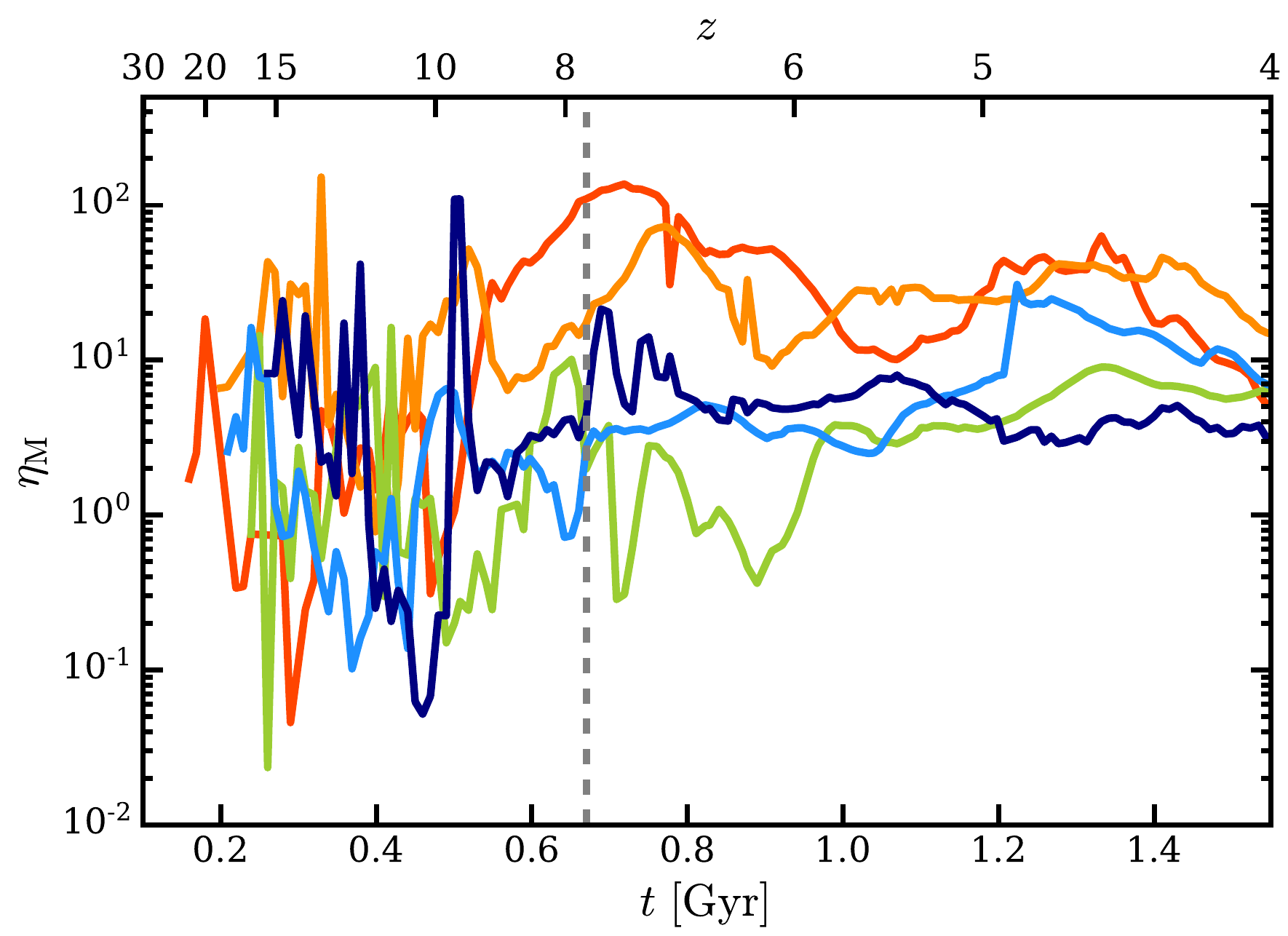}
    \caption{Mass loading factor as a function of time, using the mean SFR until $z=4$ as opposed to the instantaneous SFR.}
    \label{fig:massloading}
\end{figure}

HaloA-D were chosen to have $z=0$ halo masses within a factor of two of each other. Nevertheless, their growth histories vary. In Fig.~\ref{fig:T200} we show the virial temperature of each halo as a function of redshift:
\begin{equation}
    T_{200} = \frac{\mu  m_p}{2 k_\mathrm{B}} \frac{G  M_{200}}{R_{200}},
\end{equation}
where 
%$\mu$ is the mean molecular weight, $m_p$ is the proton mass, $k_\mathrm{B}$ is the Boltzmann constant, $G$ is the gravitational constant, % (alrady defined earlier)
$M_{200}$ is the viral mass and $R_{200}$ is the virial radius.
To guide the eye, we have also plotted the atomic cooling threshold of $10^4$ K and the quenching redshift of $z=7.7$. The magenta line shows the IGM temperature, computed following \cite{Puchwein2015}, namely the median gas temperature of IGM gas within 5\% of the mean cosmic density. The IGM temperature begins to rise at $z=8.3$ according to our UVB, and reaches $10^4~\mathrm{K}$ by $z=7.3$. Starting at $z=4$ we begin to see HeII reionization heating the IGM. After $z=2$ the IGM temperature drops below $10^4~\mathrm{K}$ due to the expansion of the universe.

We see that HaloD displays the fastest early growth, rising above $10^4$ K as early as $z\sim10$. This is also seen in the concentration parameter (see Tab.~\ref{tab:general}). The concentration is computed following \cite{Bullock2001} assuming an NFW profile shape \citep{NFW} and using the $r_\mathrm{max}$--method:
\begin{equation}
    c = 2.16 \frac{\Rvir}{r_\mathrm{max}},
\end{equation}
where $r_\mathrm{max}$ is the radius at the peak of the rotation curve, $v_\mathrm{max}$. Large concentration parameters are known to be indicative of early growth \citep[e.g.][]{Croton2007}, since the central halo builds up early, when the Universe was denser. The fact that this affects halo properties, such as SF, is known as assembly bias.
HaloA also grows sufficiently to pass this crucial threshold before the onset of reionization, despite having a much lower concentration. HaloB and HaloC cross this threshold after reionization, while HaloE never grows sufficiently. 

\begin{figure}
    \centering
    \includegraphics[width=\columnwidth]{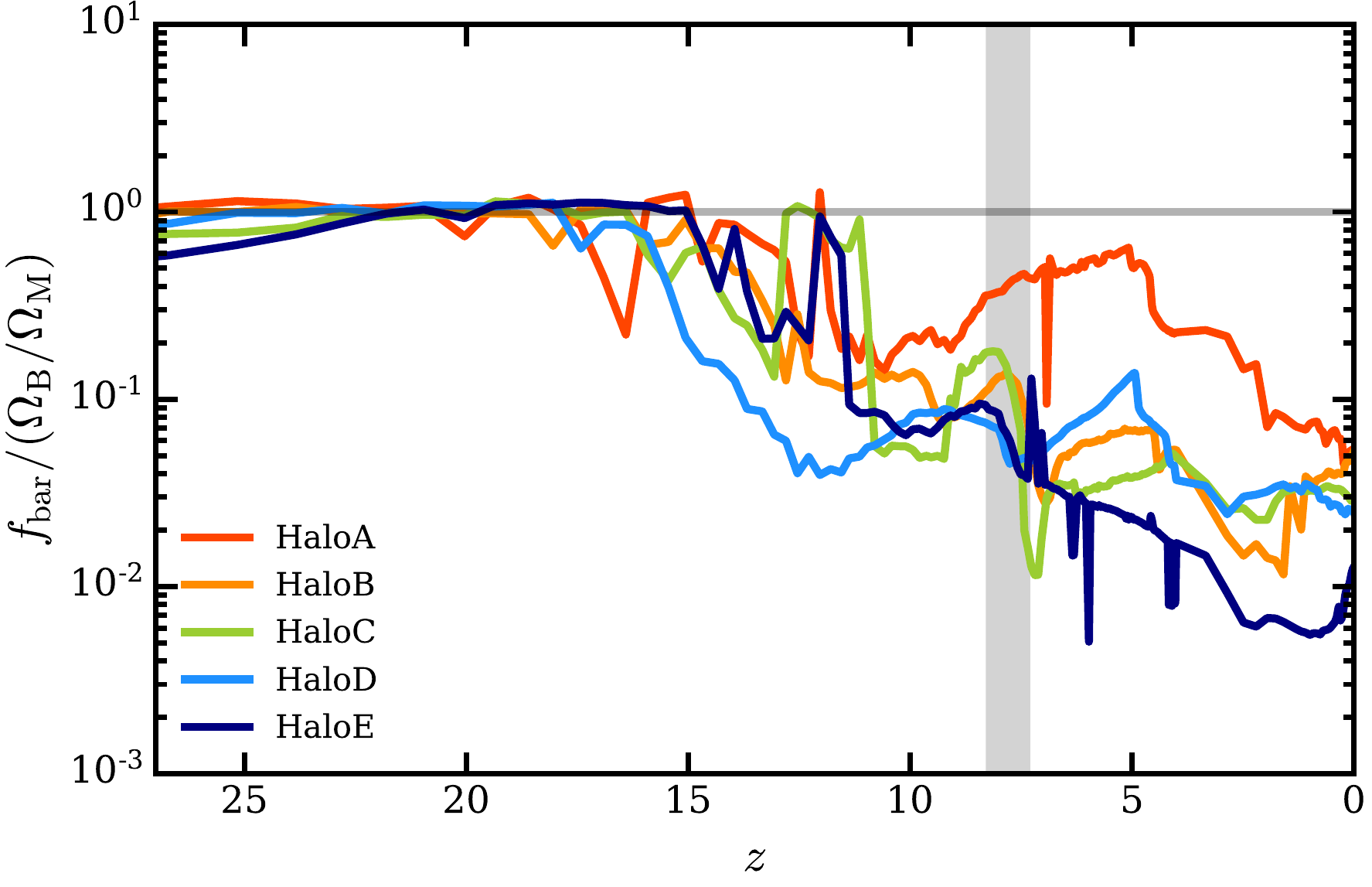}
    \caption{Baryon fraction normalized to the cosmic baryon fraction, $\Omega_\mathrm{b}/\Omega_\mathrm{m}$, as a function of redshift for each simulation. The vertical grey band indicates the time of hydrogen reionization.}
    \label{fig:baryon_fraction}
\end{figure}

Fig.~\ref{fig:massloading} shows the mass loading factor as a function of redshift for all five simulations. The mass loading factor is the ratio of the outflow rate over the mean SFR, where we have measured the mean value from the time of the first star to $z=4$:
\begin{equation}
    \eta_M = \frac{\dot M_\mathrm{out}}{\langle \dot M_\star \rangle}.
\end{equation}
Since all simulations have discontinuous SFHs, the mean SFR is lower than if we had only accounted for times of ongoing SF. Therefore, the mass loading factors shown are lower than the instantaneous values would be. The time--averaged mass loading factors from the time the first star forms to $z=4$ for the simulations in alphabetical order are: $33.7$, $26.3$, $3.7$, $6.5$, and $7.0$. It is interesting to note that the mean values are above unity in all cases. This means the halos are losing substantial gas mass throughout their evolution. We show this even more clearly in Fig.~\ref{fig:baryon_fraction}, in terms of the baryon fraction across redshift. Until $z\approx15$, all halos are consistent with hosting baryons at the cosmic baryon fraction (grey dashed line). Once SF sets in, mass loss occurs. While certain halos are able to regain higher fractions for some time, the final values tend to assume extremely low values of $1-10\%$ of the mean cosmic value. It is of interest to note that HaloE does not form new stars after $z\approx7.7$ but nevertheless shows a mass loading above unity and a decreasing baryon fraction. This must therefore be attributed to photo-evaporation due to the UVB. 

\subsection{What are the z=0 properties?} \label{sec:redshift0}

\begin{figure*}[t]
    \centering
    \includegraphics[width=\textwidth]{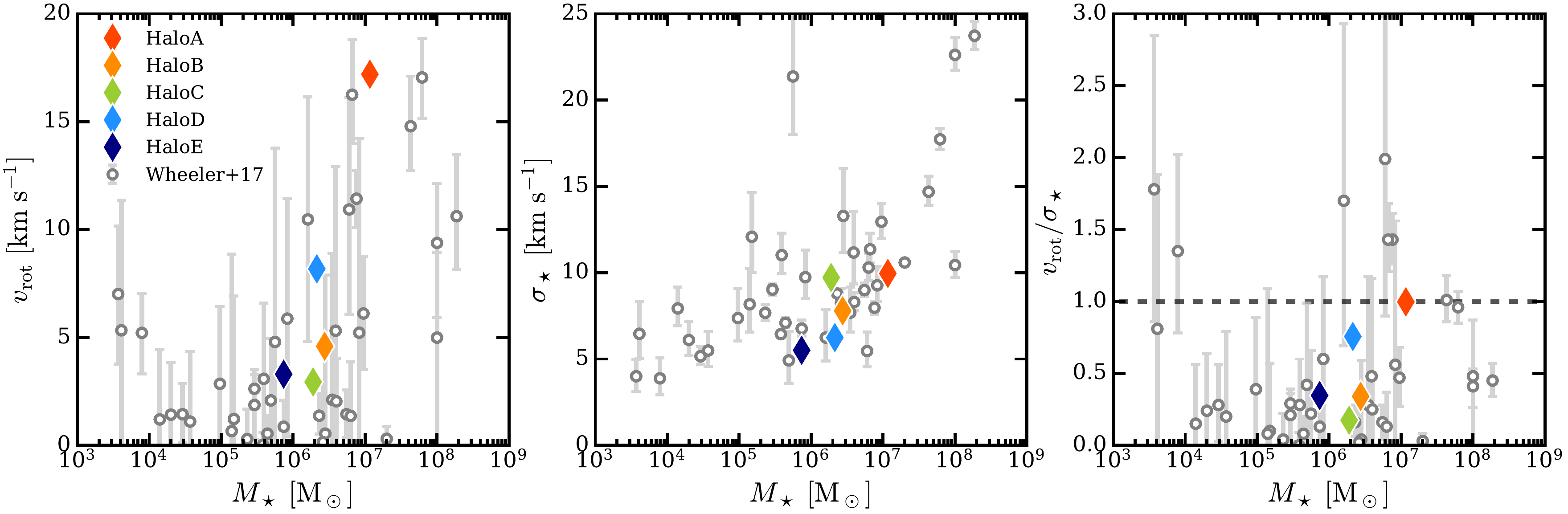}
    \caption{The peak of the stellar rotation curve, $v_\mathrm{rot}$, the 1D-velocity dispersion within the half-mass radius, $\sigma_\star$ and their ratio as a function of total stellar mass. Grey data points with error bars show the observations from \cite{Wheeler2017}. HaloA and HaloD are rotationally supported.}
    \label{fig:kinematics}
\end{figure*}

Now that we have discussed the SFHs and the reason why certain halos survive reionization, let us turn to the $z=0$ properties. Do these properties conform to Local Group observations? Do they betray their different histories? 

Firstly, in Fig.~\ref{fig:kinematics} we look at the kinematics of the stars at $z=0$. In the left panel, we show the peak value of the stellar rotation curve. The central panel is the stellar velocity dispersion measured within the half--mass radius. To be precise, this is the average of the three independent ($x$, $y$ and $z$) line-of-sight velocity dispersions. The right panel shows the ratio of these two values, giving us an indication of the dominant mode of support against gravitational collapse. In each panel, the grey data points are observations presented in \cite{Wheeler2017}. The rotation velocity of our five halos as a function of total stellar mass matches the data well. There is a distinct increase in velocity with increasing stellar mass which reflects the same trend seen in the data. Also, the velocity dispersions in the central panel match the data. However, they seem to be slightly low in the simulations. This is likely caused by the ``softened'' treatment of gravitational N--body forces below the softening length, $\varepsilon_\star$, which we have set to $4\,{\rm pc}$ in this work. This means that the gravitational acceleration acting on stars passing within $\varepsilon_\star$ of another particle is reduced, a choice made for computational efficiency. This can affect the estimated value of the velocity dispersion when interactions such as encounters closer than the softening length and binary stars interacting with a third star occur. Such interactions would cause an exchange of velocities, and, therefore, increase the total dispersion of the system. Since they are not accounted for here, the velocity dispersion is on the low end of what is expected from observations. Finally, in the right panel we see that the four lower mass systems are dispersion-dominated (values below unity). HaloA, on the other hand, sits close to unity, consistent with the appearance of more rotation-dominated systems at these stellar masses in the observational data.

\begin{figure}
    \centering
    \includegraphics[width=\columnwidth]{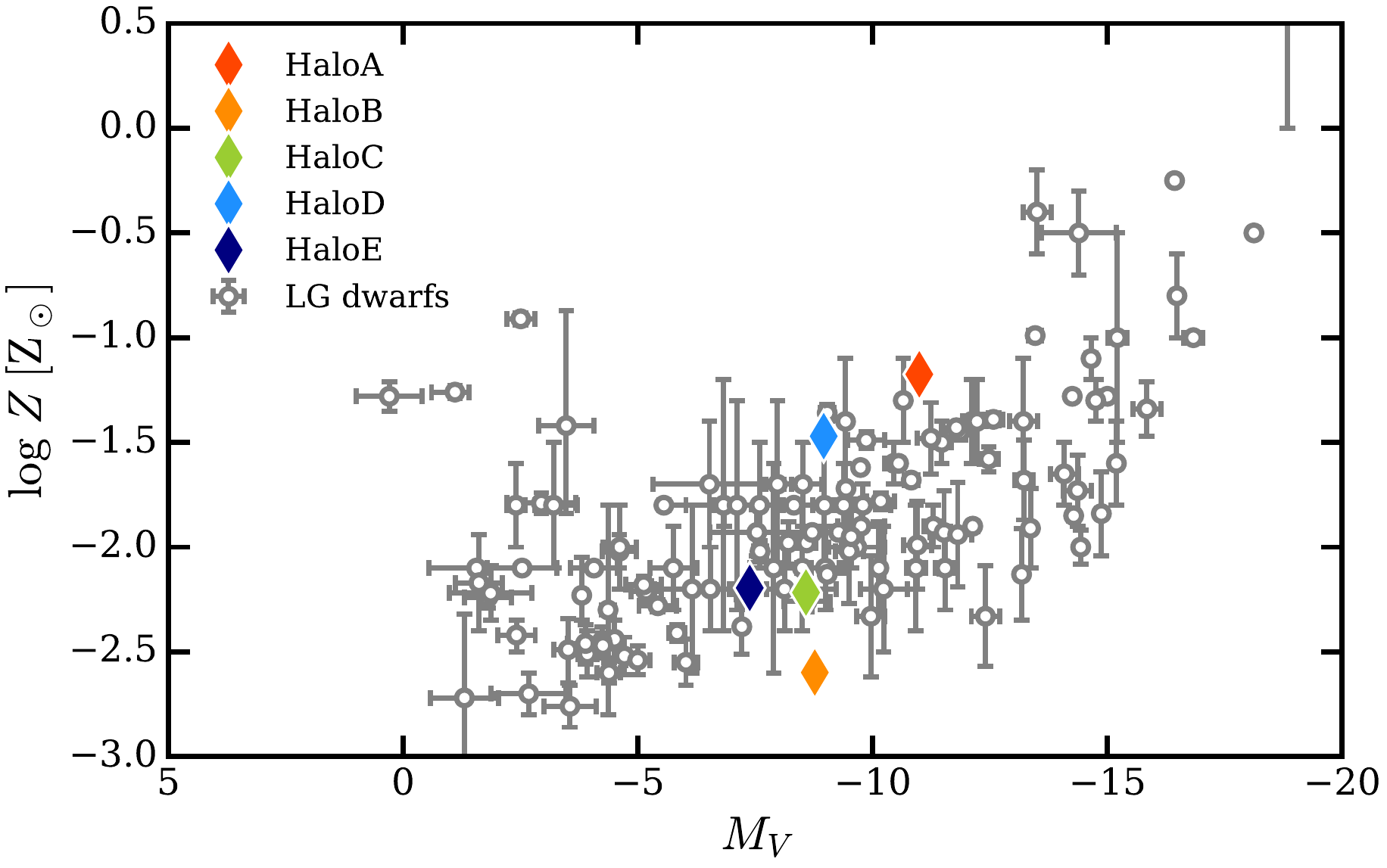}
    \caption{Metallicity as a function of V-band magnitude. Grey data points show LG dwarf observations from \cite{McConnachie2012}.}
    \label{fig:metallicity}
\end{figure}

\begin{figure}
    \centering
    \includegraphics[width=\columnwidth]{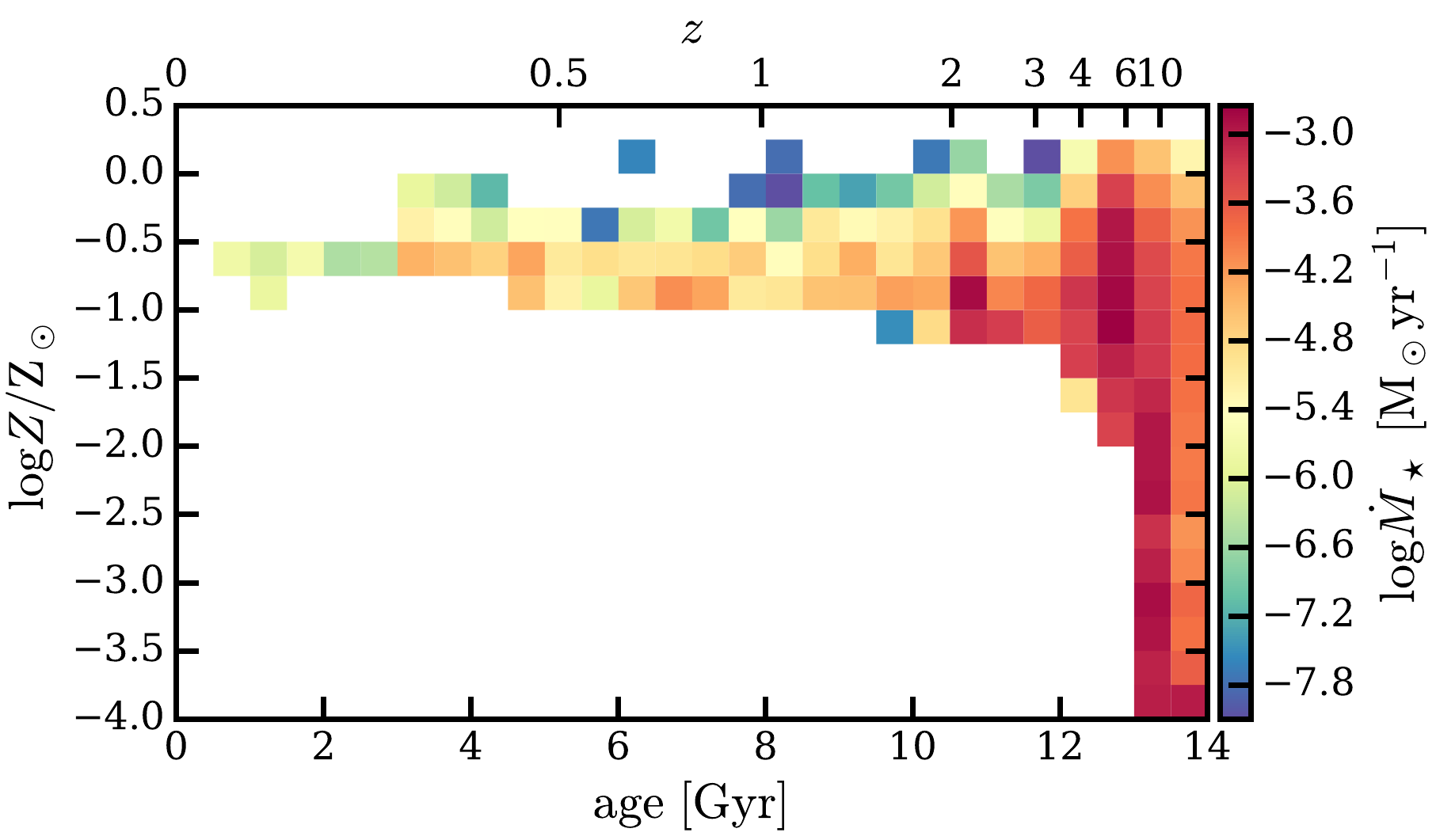}
    \caption{Age--metallicity relation for HaloA. The colorbar shows the logarithm of the mean SFR in each bin of age and metallicity. Age bins are 500~Myr wide, while metallicity bins are 0.25~dex. Compare this figure to Fig.~8a of \cite{Savino2019} showing the age-metallicity relation of Tucana dSph.}
    \label{fig:age}
\end{figure}

Fig.~\ref{fig:metallicity} shows the magnitude--metallicity relation for our simulations at $z=0$ and the LG data from \cite{McConnachie2012}. The metallicity is the log--averaged $z=0$ metallicity of all stars within \Rvir. Our simulations span the range of metallicities in the data, between $\log Z/\Zsun = -2.5$ to $-1$. As \cite{Agertz2020} showed, the metallicity relation is a useful indicator for the strength of feedback. Metallicities above the relation indicate a too weak feedback prescription, since metals are not ejected out of the ISM, instead accumulating and forming high metallicity stars. On the other hand, if the feedback is too strong then metals are preferentially ejected with the SN bursts and impoverish the ISM. The fact that this work spans the data is an encouraging indication that resolving the cooling radius for each SN self-consistently calibrates the strength of feedback. 

It is unsurprising that the two halos that survive reionization also show the highest metallicities, since later SF produces more enriched stars. To show this in detail, we plot the age--metallicity diagram for HaloA in Fig.~\ref{fig:age}. SF starts before reionization at the lowest metallicity allowed for SF, $\log Z/\Zsun = -4$. Within the first Gyr, the metallicity increases to $>-2$. After $z=4$, no new stars form below $\log Z/\Zsun = -1.5$. However, the mean SFR is also reduced significantly.
Recent Early Release Observations with the James Webb Space Telescope (JWST) using the Near-Infrared Spectrograph (NIRSpec) indicate a rapid build-up of metallicity at $z>7$ \citep{Arellano-Cordova2022}. The metallicity build-up within our halos prior to the onset of reionization is also rapid, with the first solar metallicity stars forming as early as $z>10$.

\begin{figure}
    \centering
    \includegraphics[width=\columnwidth]{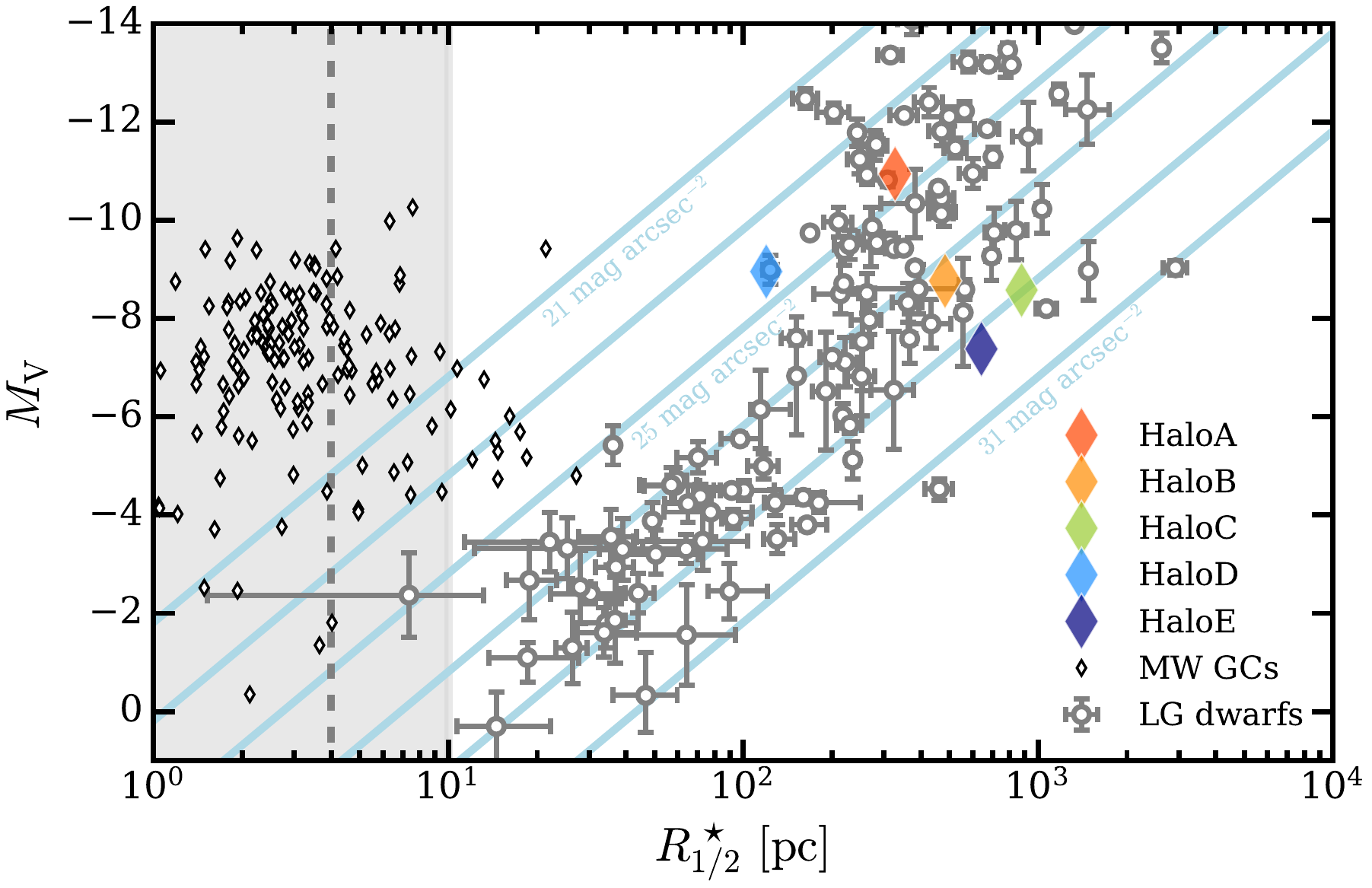}
    \caption{V-band magnitude as a function of stellar half mass (light) radius, $R_{1/2}$. Grey data points show LG dwarf observations from \cite{McConnachie2012}. Filled diamonds shows half mass radii. Small back diamonds also show the Milky Way globular clusters from \cite{Harris1996}. The grey shaded area are sizes below the dark matter softening length of our simulations. The diagonal light blue lines are lines of constant surface brightness.}
    \label{fig:rhalf}
\end{figure}

\begin{figure}
    \centering
    \includegraphics[width=\columnwidth]{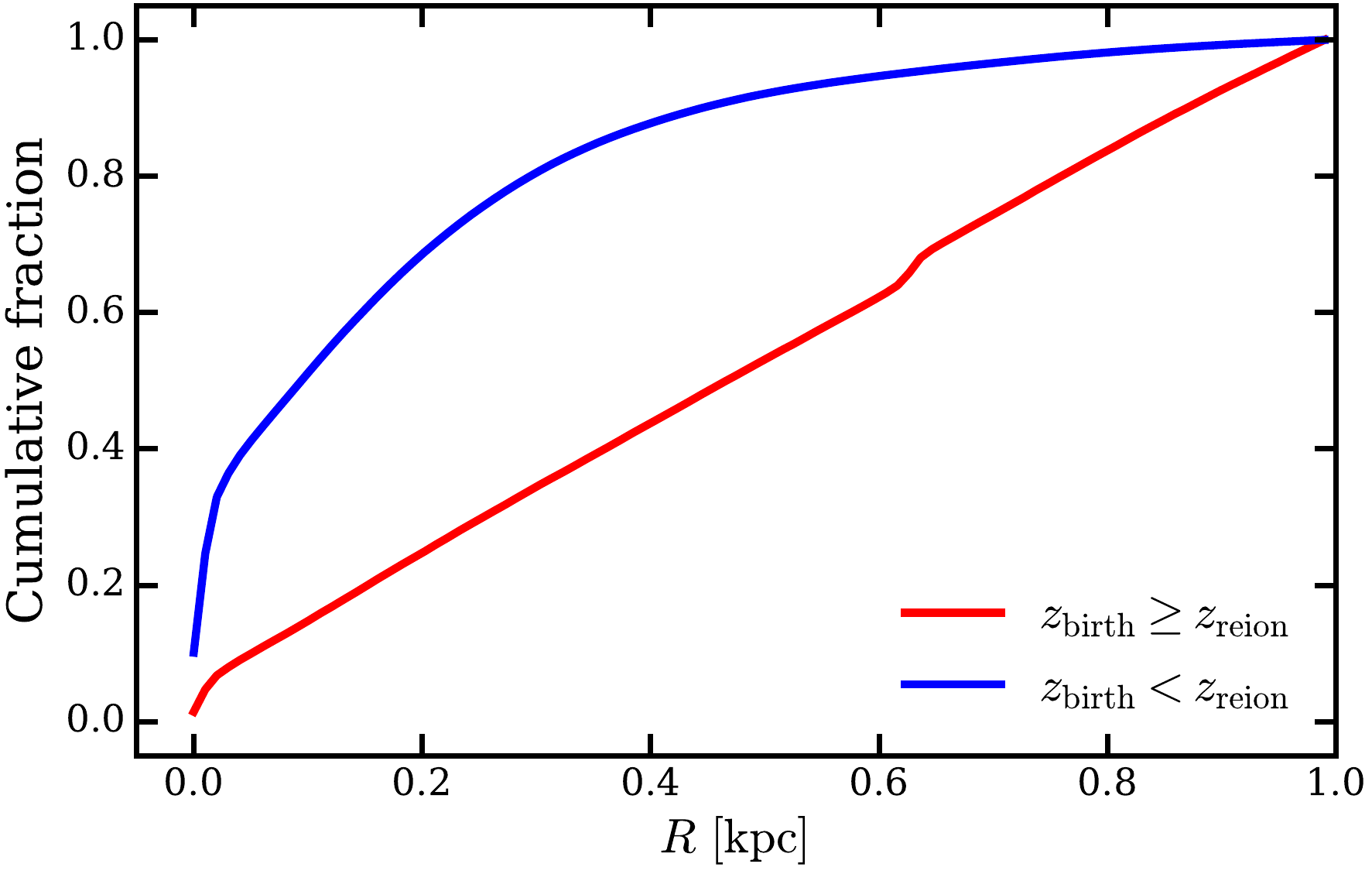}
    \caption{Cumulative radial distribution of stars within $R=1~\mathrm{kpc}$ for HaloA. The red line shows the distribution for stars formed before reionization, while the blue line is all stars formed afterwards. Clearly, the younger stars are more centrally concentrated.}
    \label{fig:profile}
\end{figure}

\begin{figure*}
    \centering
    \includegraphics[width=\textwidth]{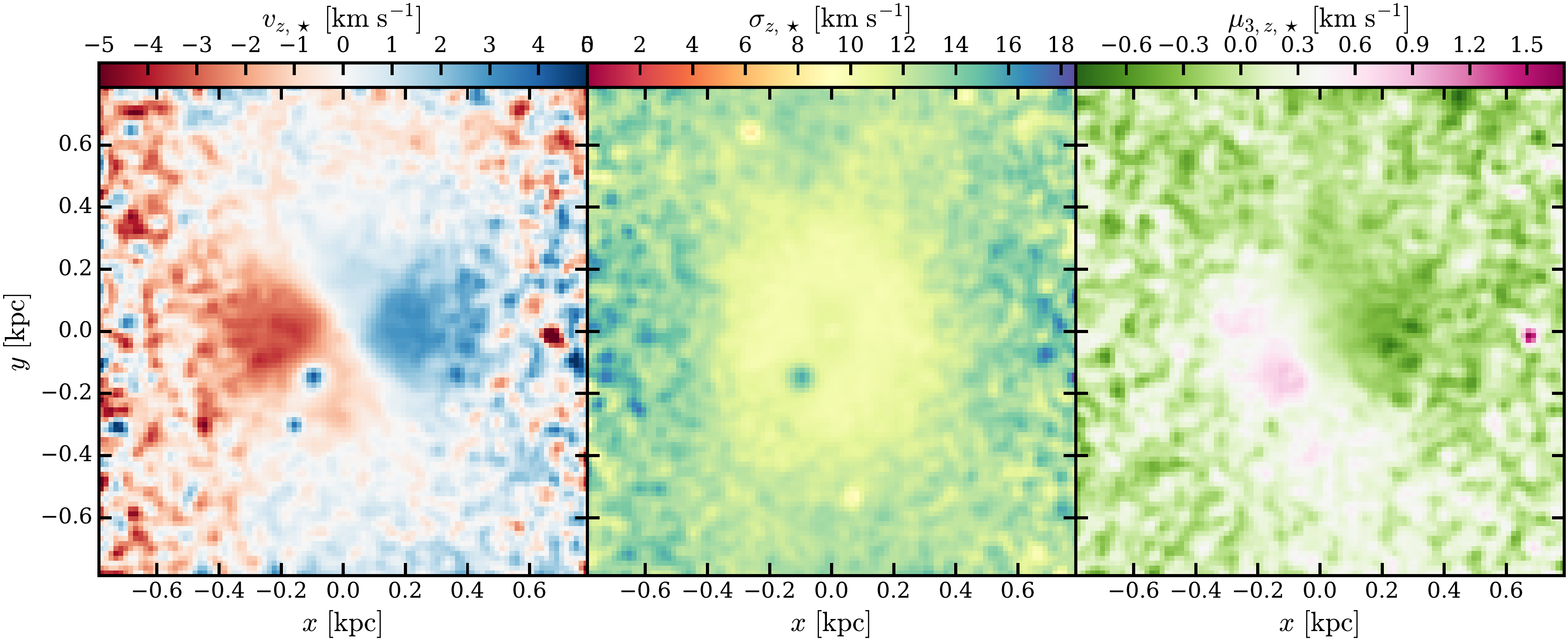}
    \caption{Line-of-sight ($z$-axis) stellar velocity (left), velocity dispersion (center) and skewness (right) of the central $5R^\star_{1/2}$ for HaloA. The halo was rotated such that the angular momentum vector of the stars points in the $y$-direction, thus we see a rotating disk edge-on. Some stellar substructure is visible.}
    \label{fig:velocity}
\end{figure*}

Finally, we also show the size--magnitude relation in Fig.~\ref{fig:rhalf}. Colored diamonds give the stellar half--mass radii in the simulations.  Again, the grey data points show LG dwarfs from \cite{McConnachie2012}. For comparison, black diamonds show Milky Way globular clusters (GCs) from \citet[2010 edition]{Harris1996}. Clearly, the simulated sizes are within expectations of the data. All five simulations are well placed within the classical dwarf regime, not in the ultra-faint regime ($M_V \lesssim -6$). The two halos with more recent SF have smaller sizes. This is due to the fact that younger stars, which are brighter and have not lost mass, are more centrally located. We show this in Fig.~\ref{fig:profile} for HaloA. This is consistent with Fig.~11b of \cite{Savino2019} for Tucana dSph. In Fig.~\ref{fig:velocity} we see that the central 0.5~kpc of HaloA are also dominated by a rotating disk, which is composed primarily of younger stars. This is line with \ref{fig:kinematics}, in which only HaloA shows signs of rotational support.

\section{Discussion} \label{sec:discussion}

Cosmological simulations at mass resolutions of $<10~\Msun$ are not yet very common. But this level of resolution is crucial for the galaxy mass scale we study here, since it allows us to resolve the multi-phase ISM and individual SN bursts without parameters tuned to match observations. We can compare our results with the EDGE simulations \citep{Agertz2020} that have a mass resolution of $20\,\Msun$. In \citet[][hereafter \citetalias{Rey2020}]{Rey2020}, the SFHs of four galaxies from this project were presented in detail. The authors show that reionization initially quenches all four. However, two are able to reignite SF at low rates after $z=1$. These two cases are reminiscent of HaloB in our sample. Of the two quenched halos in \citetalias{Rey2020}, the ``gas-rich'' simulation is similar to HaloC presented here. While neither forms stars after reionization, they are very close to doing so. 

\citetalias{Rey2020} show that a small ``genetic'' modification that increases the final mass can incite this halo to rejuvenate. In our work, HaloC is the same initial condition as presented in \cite{Gutcke2022}. In that paper, we showed that small modifications to the treatment of metals at high redshift (for example setting $M_\mathrm{PopIII} = 10^7\,\Msun$) also had the effect of inducing rejuvenation. Thus, both of these halos seem to demonstrate properties at the extreme edge between survival and quenching. Minor model variations are able to tip the evolution in one direction or another. Both halos have a very similar $z=0$ total masses of $2.5\times10^9~\Msun$ \citepalias{Rey2020} and $2.16\times10^9~\Msun$ (HaloC), respectively. One additional point to note, however, is that the stellar mass of the two are almost a factor of two different: $5.9\times10^5~\Msun$ \citepalias{Rey2020} and $1.05\times10^6\Msun$ (HaloC). 

\cite{Wheeler2019} present FIRE dwarfs with a mass resolution of $30\,\Msun$. They show consistency with \cite{Onorbe2015} in which star formation in ultra-faint dwarf galaxies is quenched by reionization, but residual self-shielded gas allows stars to continue forming until $z\sim2.5$. The lack of fresh accretion, which is caused by reionization, then prevents SF after $z=2$. They also predict that all simulations producing classical dwarfs continue to form stars until at least $z=0.5$. 

These results are somewhat in conflict with our work here. This is primarily caused by quite different stellar masses in the same halo mass ranges. For example, the ``$m09_{30}$'' simulation in \cite{Wheeler2019} has a halo mass of $\Mhalo=2.5\times10^9~\Msun$, but only produces $1.2\times10^4~\Msun$ in stars. For the same halo mass, this is two orders of magnitude lower than the stellar mass formed in our work here. This discrepancy is already set before the onset of reionzation, since the SFR in our work is $\gtrsim10^{-2}~\Msun~\yr^{-1}$, and all five galaxies form $\gtrsim10^5~\Msun$ before $z=10$. ``$m09_{30}$'' on the other hand forms stars at a rate of $\lesssim10^{-4}~\Msun~\yr^{-1}$ before reionization, producing an order of magnitude less stars by the same time. 

It is plausible that these large differences so early on in the evolution are caused by a combination of two model choices. For one, the FIRE model does not account for PopIII enrichment. Possibly more importantly, the stellar feedback prescription is known to be very strong and bursty in the FIRE model. It is likely that this suppresses early SF too much in their model. This explanation is consistent with the extremely low metallicities presented in the same work. Encouragingly, the updated FIRE-3 model \citep{Hopkins2022} appears to present altered predictions for precisely this halo mass range ($1-5\times10^9~\Msun$). Fig.~9 of that paper shows stellar masses one order of magnitude higher in the new model, bringing their results in much closer agreement to ours. The remaining differences can be attributed to the additional feedback prescriptions included in FIRE-3, such as photo-ionization in HII regions, photo-electric heating, and cosmic ray heating. Future work on the LYRA model will include these and then a more conclusive comparison can be made.

\section{Conclusions} \label{sec:conclusion}

We have carried out five ultra high resolution cosmological zoom-in simulations of dwarf galaxies in the mass range $1-4\times10^9~\Msun$. At $M_\mathrm{target}=4~\Msun$, these simulations constitute the highest resolution zoom-in simulations run fully cosmologically to $z=0$ to date. Additionally, they include a resolved multi-phase ISM and individually resolved SNe explosions. As our galaxies were selected to form in isolation and distanced from larger neighbors, our simulations probe universal aspects of cosmological evolution and of the effects of reionization, without being influenced by additional effects caused by environment or infall. 

The halo mass range chosen for our study covers the edge of where galaxy formation is expected. As such, it is unsurprising that only two out of the five simulations form a significant fraction of stars after reionization. To understand why these two survive while the other three remain quenched, we analyze the star formation histories and the early evolution to $z=4$ in detail. Our main results are as follows:

\begin{enumerate}[i)]
\item
 In the studied mass range, the $z=0$ halo mass is not a good predictor for the survival after reionization.
\item
 Framing the evolution and SFH of dwarfs in terms of an evolving mass threshold is useful in a general sense, but fails to capture the details of individual galaxies.
\item 
In our sample, halos that grow to $T_{200} > 10^4~\mathrm{K}$ before reionization form stars after reionization.
\item 
The two surviving halos are initially quenched by reionization for around 500~Myr, before recommencing SF.
\item
 The non-adiabatic heating caused by reionization elicits an oscillation or breathing of the entire gas within the halo. Initially, the gas expands, then slows and, if still bound, re-collapses onto the halo. This final step can be followed by a SF episode.
\item
Late-time ($z\sim0.3$) reignition of SF occurs in one of our halos. The halo growth and gas accretion leading up to the reignition are concurrent with the drop of the strength of the UVB after $z=2$.
\end{enumerate}

To place our simulations into the context of Local Group dwarf galaxy observations, we also compare them to various $z=0$ properties of our simulations. Our main findings are:
\begin{enumerate}[i)]
\item
The feedback strength resulting from resolved SNe produces a magnitude--metallicity relation as expected from observations.
\item
Younger stars, when they exist, are more centrally concentrated and in a rotating disk.
\item
Younger stars are more metal-rich, leading to a higher mean metallicity.
\item
The LYRA model predicts that the examined halo mass range produces classical dwarfs.
\end{enumerate}

Additionally, we would like to stress that this work is based on five simulations only. So, while this does constitute a small sample, certain evolutionary behavior may still be purely stochastic. Also, while most of our model parameters are either set by first principles or are directly motivated by observations instead of being calibrated for, there remains some room for modified parameter choices or assumptions. Some of the results presented here may be sensitive to the details of these choices.

Finally, our goals for the future are to expand the sample of simulations and the parameter space covered by this work, as well as to include additional physics, in particular magneto-hydrodynamics and cosmic rays. Thus we hope to present in forthcoming work refined predictions for dwarf galaxy properties as a function of halo mass and be able to either confirm or extend common extrapolations down into the low mass galaxy regime.

\section*{Acknowledgments}
We would like to thank Romain Teyssier and Eve C. Ostriker for useful discussions.
TAG acknowledges support by NASA through the NASA Hubble Fellowship grant $\#$HF2-51480 awarded by the Space Telescope Science Institute, which is operated by the Association of Universities for Research in Astronomy, Inc., for NASA, under contract NAS5-26555.
CP acknowledges support by the European Research Council under ERC-AdG grant PICOGAL-101019746.
GLB acknowledges support from the NSF (AST-2108470, XSEDE grant MCA06N030), a NASA TCAN award 80NSSC21K1053, and the Simons Foundation.
TN acknowledges support from the Deutsche Forschungsgemeinschaft (DFG, German
Research Foundation) under Germany’s Excellence Strategy –
EXC-2094 – 390783311 from the DFG Cluster of Excellence
‘ORIGINS’.
We acknowledge the computing time provided by the Leibniz Rechenzentrum (LRZ) of the Bayrische Akademie der Wissenschaften on the machine SuperMUC-NG (pn73we).
This research was also carried out on the High Performance Computing resources of the FREYA and COBRA clusters at the Max Planck Computing and Data Facility (MPCDF, \url{https://www.mpcdf.mpg.de}) in Garching operated by the Max Planck Society (MPG).


\bibliography{bib}{}
\bibliographystyle{aasjournal}


\end{document}